\documentclass{report}

\usepackage{graphicx} 
\usepackage{multirow}
\usepackage{array} 
\usepackage{natbib}

\begin{document}
\def\bea{\begin{eqnarray}}
\def\eea{\end{eqnarray}}
\def\deffn#1{{\bf#1}}
\arraycolsep=1pt

\def\ev{{\,\rm eV}}
\def\kev{{\,\rm keV}}
\def\mev{{\,\rm MeV}}
\def\gev{{\,\rm GeV}}
\def\Hz{{\,\rm Hz}}\def\MHz{{\,\rm MHz}}\def\GHz{{\,\rm GHz}}\def\THz{{\,\rm THz}}
\def\m{{\,\rm m}}\def\cm{{\,\rm cm}}\def\nm{{\,\rm nm}}
\def\N{\,{\rm N}}
\def\gm{{\,\rm gm}}\def\kg{{\,\rm kg}}
\def\T{\,{\rm T}}
\def\J{\,{\rm J}}
\def\pc{\,{\rm pc}}
\def\kpc{{\rm\,kpc}}
\def\Mpc{{\rm\,Mpc}}
\def\msun{\,{\cal M}_\odot}
\def\kms{\,{\rm km}\sec^{-1}}
\def\yr{{\,\rm yr}}
\def\Myr{{\rm\,Myr}}
\def\Gyr{{\rm\,Gyr}}
\def\del{\vec\nabla}
\def\Ht{H$_2$}
\def\twelveco{$^{12}\hbox{C}\hbox{O}$}
\def\thirteenco{$^{13}\hbox{C}\hbox{O}$}
\def\msun{\,{\rm M}_\odot}
\def\muB{\mu_{\rm B}}\def\muN{\mu_{\rm p}}
\def\rhs{\hbox{\caps rhs}}
\def\lhs{\hbox{\caps lhs}}
\def\rms{{\caps rms}}
\def\mcmc{{\caps mcmc}}
\def\df{{\caps df}}
\def\Omegap{\Omega_{\rm p}}
\def\Rg{R_{\rm g}}
\def\cF{{\cal F}}\def\cK{{\cal K}}\def\cJ{{\cal J}}
\newcommand {\hi} {{\rm H}\,{\small\rm I}}
\def\d{{\rm d}}\def\e{{\rm e}}
\def\i{\relax\ifmmode{\rm i}\else\char16\fi}
\def\p{\partial}
\def\deg{^\circ}

{\newif\ifnotend
\notendtrue
\def\veclist{ABCDEFGHIJKLMNOPQRSTUVWXYZabcdefghijklmnopqrstuvwxyz.}
\def\top#1#2.{#1}
\def\tail#1#2.{#2.}
\loop\expandafter\xdef\csname v\expandafter\top\veclist\endcsname%
{{\noexpand\bf\expandafter\top\veclist}}
\edef\veclist{\expandafter\tail\veclist}
\if\veclist.\notendfalse\fi\ifnotend\repeat}

%
%
\def\real{\Re\hbox{{\eightrm e}}}                   
\def\imag{\Im\hbox{{\eightrm m}}}                   
%
%
\def\fracj#1#2{{\textstyle{#1\over#2}}}
%
\def\spose#1{\hbox to 0pt{#1\hss}}
\def\lta{\mathrel{\spose{\lower 3pt\hbox{$\mathchar"218$}}
     \raise 2.0pt\hbox{$\mathchar"13C$}}}
\def\gta{\mathrel{\spose{\lower 3pt\hbox{$\mathchar"218$}}
     \raise 2.0pt\hbox{$\mathchar"13E$}}}

%
\font\gkvecten=cmmib10
\font\gkvecseven=cmmib7
\let\boldgrk=\gkvecten
\let\boldgrksc=\gkvecseven

\def\gkthing#1{{\mathchoice%
	{\hbox{{\boldgrk\char#1}}}
	{\hbox{{\boldgrk\char#1}}}
	{\hbox{{\boldgrksc\char#1}}}
	{\hbox{{\boldgrksc\char#1}}}}}

\def\valpha{\gkthing{11}}
\def\vbeta{\gkthing{12}}
\def\vgamma{\gkthing{13}}
\def\vdelta{\gkthing{14}}
\def\vepsilon{\gkthing{15}}
\def\vzeta{\gkthing{16}}
\def\veta{\gkthing{17}}
\def\vtheta{\gkthing{18}}
\def\viotaeta{\gkthing{19}}
\def\vkappa{\gkthing{20}}
\def\vlambda{\gkthing{21}}
\def\vmu{\gkthing{22}}
\def\vnu{\gkthing{23}}
\def\vxi{\gkthing{24}}
\def\vpi{\gkthing{25}}
\def\vrho{\gkthing{26}}
\def\vsigma{\gkthing{27}}
\def\vtau{\gkthing{28}}
\def\vupsilon{\gkthing{29}}
\def\vphi{\gkthing{30}}
\def\vchi{\gkthing{31}}
\def\vpsi{\gkthing{32}}
\def\vomega{\gkthing{33}}

\def\vnabla{{\mathchoice
{\hbox{{\boldsym\char114}}}
{\hbox{{\boldsym\char114}}}
{\hbox{{\boldsymsc\char114}}}
{\hbox{{\boldsymsc\char114}}}
}}%

\def\vDelta{{\bf\Delta}}
\def\vOmega{{\bf\Omega}}

\def\figref#1{Fig.~\ref{#1}}

\def\Phieff{\Phi_{\rm eff}}

\def\const{\hbox{constant}}

\renewcommand{\[}{\begin{equation}}\renewcommand{\]}{\end{equation}}



\setcounter{chapter}{1}
\pagestyle{myheadings}
\markright{James Binney\hfil Dynamics of secular evolution\hfil}

\title{Dynamics of Secular Evolution}

\author{James Binney\\ (Oxford University)\\  
        Rudolf Peierls Centre for Theoretical Physics,\\
	1 Keble Road, Oxford, OX1 3NP, UK\\
	(To appear in XXIII Canary islands winter school of astrophysics\\
	ed.\ J. Falcon-Barroso \& J.H. Knapen)}

\maketitle

\abstract{The material in this article was presented in five hours of
lectures to the 2011 Tenerife Winter School. The School's theme was ``Secular
Evolution of Galaxies'' and my task was to present the underlying
stellar-dynamical theory. Other lecturers were speaking on the role of bars
and chemical evolution, so these topics are avoided here. The material
starts with an account of the connections between isolating integrals,
quasiperiodicity and angle-action variables -- these variables played a
prominent and unifying role throughout the lectures. This leads on to the
phenomenon of resonant trapping and how this can lead to chaos in cuspy
potentials and phase-space mixing in slowly evolving potentials.  Surfaces of
section and frequency analysis are introduced as diagnostics of phase-space
structure.  Real galactic potentials include a fluctuating part that drives
the system towards unattainable thermal equilibrium. Two-body encounters are
only one source of fluctuations, and all fluctuations will drive similar
evolution.  The orbit-averaged Fokker-Planck equation is derived, as are
relations that hold between the second-order diffusion coefficients and both
the power spectrum of the fluctuations and the first-order diffusion
coefficients.  From the observed heating of the solar neighbourhood we
show that the second-order diffusion coefficients must scale as $\sim
J^{1/2}$. We show that periodic spiral structure shifts angular momentum
outwards, heating at the Lindblad resonances and mixing at corotation. The
equation that would yield the normal modes of a stellar disc is first derived
and then used to discuss the propagation of tightly-wound spiral waves. The
winding up of such waves is described and explains why cool stellar discs are
responsive systems that amplify ambient noise. An explanation is offered of
why the Lin-Shu-Kalnajs dispersion relation and even global normal-mode calculations
provide a very incomplete understanding of the dynamics of stellar
discs.}

\tableofcontents

\section{Orbits}\label{sec:orbits}
 Even a gas-free galaxy is a formidably complex dynamical system:
$\sim10^{11}$ stars and inconceivably more WIMPS are strongly coupled by their
mutual gravitational field. As in any branch of theoretical physics we make
progress by simplifying. The key simplification is that the motion of all
these particles would be very similar if calculated in the gravitational
field of an imaginary smoothed-out version of the galaxy. That is, we smear
the mass of each particle over something like an interstellar distance and
compute the gravitational field from this smooth mass distribution. Given the
usefulness of this approximate gravitational field, our first task becomes to
understand how particles move in smooth gravitational fields of this kind.

\subsection{Quasiperiodicity and isolating integrals}\label{sec:quasiper}

\begin{figure}
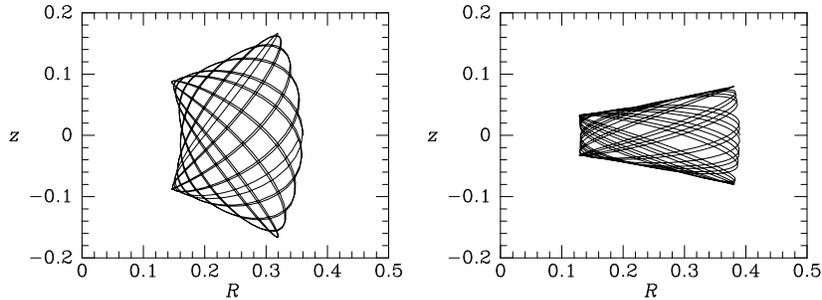

\centerline{\includegraphics[width=.43\hsize]{chapter_binney/fig1a.ps}\quad
\includegraphics[width=.43\hsize]{chapter_binney/fig1b.ps}}
\caption{The motion  in the $(R,z)$ plane of a particle that orbits in an
axisymmetric gravitational potential. The orbits have the same values of $E$
and $L_z$ but are nevertheless very different because they have
different values of the mysterious ``third integral''.   }\label{fig:one}
\end{figure}

The equations of motion of a particle in a typical gravitational potential
$\Phi$
are readily integrated numerically. If the potential is axisymmetric, the
component of angular momentum about the galaxy's symmetry axis, $L_z$, is
conserved, and we can eliminate the azimuthal angle $\phi$ and its time
derivative from the equations of motion, leaving us with two coupled ordinary
differential equations to integrate:
 \[
\ddot R=-{\p\Phieff\over\p R}\qquad
\ddot z=-{\p\Phieff\over\p z},
\]
 where
\[\label{eq:defsPhieff}
\Phieff(R,z)\equiv\Phi(R,z)+{L_z^2\over2R^2}.
\]
 \figref{fig:one} shows a couple of typical orbits obtained in this way. They
have a nicely regular pattern of paths running to and fro. Given sufficient
time such an orbit will carry the particle to any given point inside the
bounding envelope, and when it reaches that point the particle will be moving
with one of four velocities: $\dot R=\pm v_R$ and $\dot z=\pm v_z$, where
$v_R$ and $v_z$ are numbers that depend on the orbit and the location.

If the time series $R(t)$ or $z(t)$ obtained in this way is Fourier
transformed, one finds that the spectrum consists of
discrete lines, and moreover, any frequency appearing in the spectrum can be
expressed as an integer linear combination of two \deffn{fundamental
frequencies} $\Omega_R$ and $\Omega_z$. That is, 
 \begin{eqnarray}\label{eq:Roft}
R(t)=\sum_\vn R_\vn\cos(\vn\cdot\vOmega t+\psi_{R\vn}),\qquad
z(t)=\sum_\vn Z_\vn\cos(\vn\cdot\vOmega t+\psi_{z\vn}),
\end{eqnarray}
 where
\[
\vOmega=(\Omega_R,\Omega_z)\quad\hbox{and}\quad
\vn=(n_R,n_z)
\]
 with $n_R$ and $n_z$ being (possibly negative) integers. Thus the orbit is
characterised by the two fundamental frequencies and countably many
amplitudes $R_\vn,z_\vn$ and phases $\psi_{R\vn},\psi_{z\vn}$. An orbit with
these properties is said to be \deffn{quasiperiodic}. 

As the particle moves
in the $(R,z)$ plane, it rotates about the symmetry axis: its azimuthal
coordinate $\phi$ can be found by simple quadrature:
 \[
\phi(t)=\phi(0)+\int_0^t\d t\,{L_z\over R^2}.
\]
 The time series of $\phi$ will consist of sinusoidal contributions at
discrete frequencies superimposed on a secularly increasing value, and the
rate of the secular increase defines a third fundamental frequency
$\Omega_\phi$. Thus three fundamental frequencies will be required to express
the frequency of any line in the spectrum of say $x(t)$ as an integer linear
combination of fundamental frequencies.

The result of integrating the equations of motion in potentials that are not
axisymmetric cannot be displayed as nicely as in \figref{fig:one}. But when
you Fourier transform the time series $x(t)$ of the coordinates of such
orbits, you usually find that the spectra consist of lines that can be
expressed as integer linear combinations of three fundamental frequencies. So
a large fraction of orbits in galaxy-like potentials are quasiperiodic.

An integral of motion is a function $I(\vx,\vv)$ on phase space that returns
the same number no matter at what point along an orbit you evaluate it:
 \[\label{eq:defI}
I(\vx(t),\vv(t))=\const.
\]
 We say that $I$ is an \deffn{isolating integral} if the set of phase-space
points that satisfy equation (\ref{eq:defI}) for some given value on the
right side defines a smooth five-dimensional subset of phase space. For
example, the energy $E(\vx,\vv)=\frac12v^2+\Phi(\vx)$ is an isolating
integral. 

If an orbit is quasiperiodic, one may show that it admits at least three
functionally independent isolating integrals. We can take $E$ to be one of
these, and in the case of an axisymmetric potential, $L_z=xv_y-yv_x$ can be
taken to be another of them. Only in exceptional potentials do we have an
analytic expression for a third isolating integral, but its existence is
assured by the quasiperiodic nature of orbits -- for a proof see
\cite{Arnold}.

\subsection{Angles and actions}\label{sec:angleact}

It often happens that the key to solving a physics problem is to identify the
coordinate system that's best suited to that problem.  Hamiltonian mechanics
allows us to use an extremely wide range of coordinates for phase space with
ease -- all ``canonical coordinates'' are intrinsically equal. In the
following we shall denote by $(\vx,\vp)$ a canonical system made up of
coordinate $x_i$ that gives the star's position and its canonically conjugate
momentum $p_i$. In the simplest case $x_i$ is a Cartesian coordinate and
$p_i=\dot x_i$ is its rate of change, but in some instances $p_i\ne \dot
x_i$.  Given that three isolating integrals $I_i$ exist, a shrewd question to
ask is ``is there a canonical coordinate system in which the $I_i$, or some
functions of them, act as momenta?'' Since any function $J(I_1,I_2,I_3)$ of
the $I_i$ will itself be an integral of motion, we usefully increase our
chances of our enquiry having the answer ``yes'' by opening the enquiry up to
functions of our original integrals.

It turns out that only a highly restricted group of integrals are capable of
playing the role of momenta. We call such integrals \deffn{actions} and
reserve for them the letter $J$, so $J_j(\vx,\vp)$ is the $j$th action of an
orbit. 

To understand why only special integrals can act as momenta, we should
consider the subset $M$ of points in phase space that satisfy
 \[\label{eq:3I}
I_i(\vx,\vp)=\const\quad\hbox{for }i=1,2,3.
\]
 These equations impose three restrictions on the six phase-space coordinates
$(\vx,\vp)$. So the set $M$ is a three-dimensional subset of phase space to
which the particle is confined for all time. $M$ is a subset of the energy
hypersurface $H(\vx,\vp)=E$, where $H$ is the Hamiltonian function and $E$ is
the particle's energy. If the orbit is bound, as we shall assume, the energy
hypersurface is compact, so $M$ is compact also. In practice $M$ will also be
a connected set, and it can then be shown that it is diffeomorphic to a
3-torus \citep[see \S\S49,50 of][for a proof]{Arnold}. 

\begin{figure}
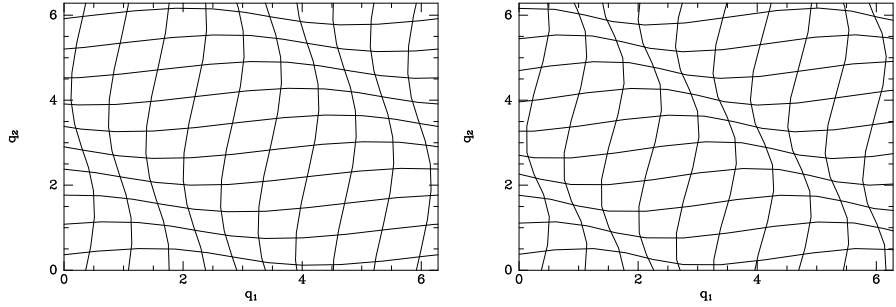

\centerline{\includegraphics[width=.47\hsize]{chapter_binney/fig2a.ps}\quad
\includegraphics[width=.47\hsize]{chapter_binney/fig2b.ps}}
\caption{Trajectories generated by an action (left) and by some other
integral (right). In the left panel each horizontal curve joins points on the left
and right boundaries that have been identified, whereas in the right panel the
leftmost and rightmost points of these curves are at logically distinct
points. Consequently, in the right panel the periodic continuation of each
horizontal curve would not overlie the portion plotted. The vertical curves
exhibit the same phenomenon. Consequently these curves define a global
coordinate system only in the left panel.}\label{fig:traj}
\end{figure}

What is a \deffn{3-torus}?  Think of a room with each point on the floor identified
with the point on the ceiling that is vertically above it, each point on the
left wall identified with the corresponding point on the right wall, and with
points on the front and back walls similarly identified. Position within this
room ($M$) is specified by the values of the coordinates $\theta_1$,
$\theta_2$ and $\theta_3$ that are canonically conjugate to $J_1$, $J_2$ and
$J_3$, whose values specify {\it which\/} room (torus) we are in. They do so
in a way that can be understood geometrically.

In Hamiltonian mechanics a fundamental role is played by the \deffn{Poincar\'e
invariant} $P$. This is a number that we assign to any two-surface $S$ in phase
space through
 \[
P(S)\equiv\sum_i\int_S\d x_i\d p_i,
\]
 so $P(S)$ is the sum of the areas of the projections of $S$ onto all the
planes formed by each coordinate $x_i$ and its conjugate momentum $p_i$. It's
called an {\it invariant\/} because you get the same number no matter what
canonical coordinates you use. Consequently, given any two-surface $S\in M$
we can replace $(x_i,p_i)$ by $(\theta_i,J_i)$ and write
 \[
P(S)=\sum_i\int_S\d\theta_i\,\d J_i
\]
 and this vanishes because every point of $S$ has the same value of $J_i$.
In view of this fact we say $M$ is \deffn{null}. It follows from the nullness
of $M$ that the value of any line integral $\int\d\vx\cdot\vp$ through $M$
between two given endpoints is the same for any path between those points
that can be continuously distorted into some standard path without leaving
$M$. To see why, take the difference between the integrals along two
different paths $\Gamma_1$ and $\Gamma_2$ between a given pair of points.
This difference of integrals is identical with the line integral along the
closed path $\Gamma_1-\Gamma_2$ in which we go out on $\Gamma_1$ and back on
$\Gamma_2$.  By Green's theorem this closed line integral is equal to the
Poincar\'e invariant of the 2-surface that $\Gamma_2$ sweeps out as it is
deformed into $\Gamma_1$. But this Poincar\'e invariant vanishes, so the
original line integrals were equal. Consider now the line integral
$\int\d\vx\cdot\vp$ along the path from the front wall to the corresponding point
on the back wall. We choose to evaluate this integral using the
$(\vtheta,\vJ)$ coordinates, and to take the path on which
$\theta_2=\theta_3=\const$. Then we have
 \[
\int\d\vx\cdot\vp=\int\d\vtheta\cdot\vJ=J_1\int\d\theta_1
\]
so $J_1$ is equal to  $\int\d\vx\cdot\vp$ divided by the amount by which
$\theta_1$ increments as we cross the room. We choose to scale the actions so
$\theta_i$ increments by $2\pi$ on crossing the room, so
\[
J_i={1\over2\pi}\oint_{\Gamma_i}\d\vx\cdot\vp
={1\over2\pi}\sum_j\oint_{\Gamma_i}\d x_jp_j,
\]
 where $\Gamma_i$ is any path that takes us once across the room.

If we want to find the values of the ordinary phase-space coordinates
$(\vx,\vp)$ at the point $\vtheta$ we need to do the following: (i) Choose a
point $(\vx_0,\vp_0)$ in the room and declare it to be $\vtheta=0$. Now
integrate from the initial conditions $\vx=\vx_0, \vp=\vp_0$ when
$\theta_1=0$ the coupled o.d.e.s 
 \[
{\d x_j\over\d\theta_1}=[x_j,J_1],\qquad
{\d p_j\over\d\theta_1}=[p_j,J_1],
\]
 where $[..]$ denotes a Poisson bracket, to discover the $(\vx,\vp)$
coordinates of the points $\vtheta=(\theta_1,0,0)$. (ii) Starting from  any of
these points we can integrate the o.d.e.s 
 \[
{\d x_j\over\d\theta_2}=[x_j,J_2],\qquad
{\d p_j\over\d\theta_2}=[p_j,J_2],
\]
 to discover the $(\vx,\vp)$ coordinates of the points
$\vtheta=(\theta_1,\theta_2,0)$. (iii) Starting from any of the
last-mentioned points we can integrate the third set of o.d.e.s to find the
$(\vx,\vp)$ coordinates of a general point $\vtheta$. The key property of
actions is that if we integrate say the first set of o.d.e.s from a point
$\vtheta_a$ on one wall to the point $\vtheta_b$ at which the trajectory hits
the opposite wall, we find that $\vtheta_b=\vtheta_a+(2\pi,0,0)$
(\figref{fig:traj} left). That is the trajectories generated by actions carry
you from a point on one wall to the point on the opposite wall that we have
identified with our starting point. If we form some other integrals
$I_i(\vJ)$ by adopting three non-trivial functions of the actions, the
trajectories generated by the $I_i$ will not carry us from a point on a wall
to the point on the opposite wall that has been identified with it
(\figref{fig:traj} right).  Consequently, the variables that are canonically
conjugate to the $I_i$ cannot form a {\it global\/} coordinate system
(although they can be used as coordinates locally).

In remembrance of the fact that $\vtheta_a$ and
$\vtheta_b=\vtheta_a+(2\pi,0,0)$ correspond to the same point in phase space,
the canonically conjugate coordinates of actions are called \deffn{angle
variables}.  To see what flexibility we have in the choice of actions and
angles, we observe that if we have new variables $\vtheta'$ that are related
to our original angle variables by linear equations
 \[\label{eq:thetaofthetap}
\vtheta=\vN\cdot\vtheta',
\]
 where $\vN$ is a matrix with integer entries, then if any of the $\theta'_i$
is incremented by $2\pi$, all the $\theta_j$ will change by $2m\pi$, so we
will return to the same place in phase space. Hence the $\vtheta'$ provide a
global coordinate system for a torus as effectively as the $\vtheta$. To
discover what actions correspond to the $\vtheta'$, we write down the
generating function of the canonical transformation
$(\vtheta,\vJ)\leftrightarrow(\vtheta',\vJ')$
 \[\label{eq:newangs}
S(\vtheta',\vJ)=\sum_{ij}J_iN_{ij}\theta'_j.
\]
 It is straightforward to check that $S$ generates equation
 (\ref{eq:thetaofthetap}) and we have also
\[
J'_j={\p S\over\p\theta'_j}=\sum_iJ_iN_{ij}.
\]
 Thus the new actions are integer linear combinations of the old actions.

Any function on phase space can be expressed as a function of the angles and
actions $(\vtheta,\vJ)$. In particular the Hamiltonian can be
expressed in this form. But while the angle variables vary as a particle
orbits, its energy does not. So the Hamiltonian cannot depend on
$\vtheta$. Hence the Hamiltonian is a function $H(\vJ)$ of the actions only.
This fact makes the equations of motion of the angle variables trivial:
 \[\label{eq:defsOmega}
\dot\theta_i={\p H\over\p J_i}\equiv\Omega_i.
\]
 Since $H$ depends on $\vJ$ only, so must its partial derivatives $\Omega_i$.
Since $\vJ$ is constant along the orbit, $\Omega_i$ is too, so we can
immediately integrate equation (\ref{eq:defsOmega}):
 \[\label{eq:thetaoft}
\theta_i(t)=\theta_i(0)+\Omega_it.
\]
 Thus the particle moves through its room at constant speed along straight
lines whose slope is given by $\vOmega=\p H/\p\vJ$.  

\begin{figure}
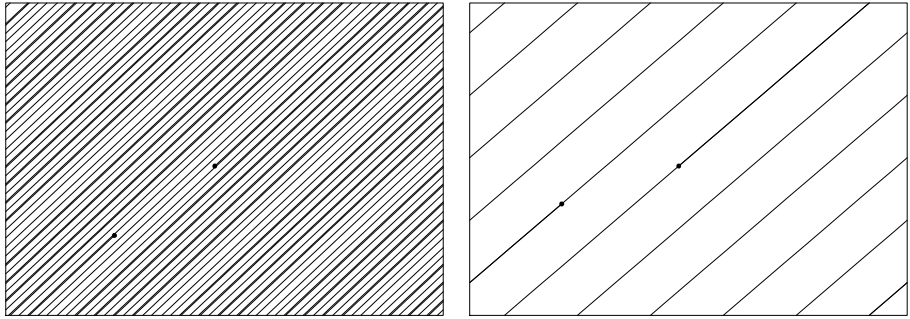

\centerline{ \includegraphics[width=.48\hsize]{chapter_binney/fig3a.ps}\quad
\includegraphics[width=.48\hsize]{chapter_binney/fig3b.ps} } \caption{A
particle moves through its room (torus) on a straight line that repeatedly
hits a wall and reappears at the corresponding point on the opposite wall. If
the components of the slope vector are rationally related, the line
eventually closes on itself (right). In general it does not close and comes
arbitrarily close to every point in the room (left). Black dots mark the
start and end of the plotted portion of the trajectory.}\label{fig:straight}
\end{figure}

Usually the frequencies are \deffn{incommensurable} and then the line will
eventually come arbitrarily close to every point in the room
(\figref{fig:straight} left).

Since the ordinary phase-space coordinates are periodic functions of the
angle variables with period $2\pi$ they can be expanded in Fourier series
 \[
x_i(\vtheta)=\sum_\vn X_{i\vn}\cos(\vn\cdot\vtheta+\psi_{i\vn}),
\]
 where the sum is over vectors with integer components and the $X_{i\vn}$ and
$\psi_{i\vn}$ are constant amplitudes and phases.
When we substitute into this expression our solution (\ref{eq:thetaoft}) of
the equations of motion, we obtain
 \[
x_i(t)=\sum_\vn X_{i\vn}\cos(\vn\cdot\vOmega\, t+\psi'_{i\vn}),
\quad \hbox{where} \quad
\psi'_{i\vn}\equiv\psi_{i\vn}+\vn\cdot\vtheta_i(0).
\]
 Equation (\ref{eq:Roft}) from which we started is an instance of this
equation. Thus we have come full circle from the empirical fact that
numerically integrated orbits are quasiperiodic to the fact that such spectra
are a necessary consequence of these orbits having three isolating
integrals.

In the generic case of incommensurable frequencies, we have a useful result,
the \deffn{time-averages theorem}: {\it when the frequencies are
incommensurable, the fraction of its time a particle spends in a subset $V$
of the torus is $\int_V\d^3\vtheta/(2\pi)^3$.} From this the \deffn{strong Jeans
theorem follows}: {\it in a steady-state galaxy, the density of stars is
uniform within incommensurable tori, so the density of stars in phase space
is a function $f(\vJ)$ of the actions only.} We call $f$ the
\deffn{distribution function}, often abbreviated \deffn{DF}.

It is useful to think of orbits as points in the three-dimensional space,
\deffn{action space}, that has the three actions as is Cartesian coordinates.
The strong Jeans theorem tells us that galactic equilibria are simply
distributions of stars in this easily imagined space.

\subsection{Epicycle approximation}\label{sec:epicycle}

It's instructive to examine a very useful model at this point. A star in an
axisymmetric potential moves in the effective potential
(\ref{eq:defsPhieff}). It is physically obvious  that the minimum of $\Phieff$
occurs at $(\Rg,0)$, where the \deffn{guiding-centre radius} $\Rg$ is the
radius of the circular orbit with the given angular momentum $L_z$. Since
this is the potential's minimum, a  Taylor expansion of $\Phieff$ around it
will contain no linear terms and be of the form
 \[
\Phieff(\Rg+x,z)=\Phieff(\Rg,0)+\fracj12\kappa^2x^2+\fracj12\nu^2z^2+\cdots
\]
 Stars on sufficiently circular orbits will be confined to the region in
which we need retain only the first three terms in this series. Consequently,
their radial and vertical motions will be harmonic. The frequencies of these
oscillations are $\Omega_r=\kappa$, the \deffn{epicycle frequency} and
$\Omega_z=\nu$, the \deffn{vertical epicycle frequency}. The solution to the
equations of motion is
\[
R(t)=\Rg+X\cos(\kappa t+\psi_r)\quad;\quad z(t)=Z\cos(\nu t+\psi_z),
\]
 where $X,\psi_r,Z$ and $\psi_z$ are all constants. Clearly we can set
\[
\theta_r=\kappa t+\psi_r\quad;\quad
\theta_z=\nu t+\psi_z.
\]
 Then calculating $p_R=\dot
R$ and evaluating $\oint\d R\,p_R$ we can show that $X=\sqrt{2J_r/\kappa}$,
and similarly that $Z=\sqrt{2J_z/\nu}$, so in the epicycle approximation the
connection between ordinary coordinates and angle-action variables is
 \[\label{eq:epiactions}
R=\Rg+\sqrt{{2J_r\over\kappa}}\cos\theta_r\quad;\quad
z=\sqrt{{2J_z\over\nu}}\cos\theta_z.
\]

\section{Resonances}\label{sec:resonance}

The fundamental frequencies $\vOmega$ are generally non-trivial functions of
the actions, so for certain tori a \deffn{resonance condition}
$\vN\cdot\Omega=0$ will apply. When we multiply the resonance condition by
$t$ and use equation (\ref{eq:thetaoft}), we obtain
 \[
\vN\cdot\vtheta(t)=\const.
\]
 This equation inspires us to make a canonical transformation to new angles
and actions $(\vtheta',\vJ')$ such that $\vN\cdot\vtheta$ becomes one of the
new angles, say $\theta'_1$.  It does not much matter what we adopt for $\theta'_2$ and
$\theta'_3$; $\theta'_2=\theta_2$ and $\theta'_3=\theta_3$ works fine. Then
our generating function becomes
\[\label{eq:resS}
S(\vtheta,\vJ')=J'_1\vN\cdot\vtheta+J'_2\theta_2+J'_3\theta_3,
\]
 so the new angles are
 \[\label{eq:restheta}
\theta'_1={\p S\over\p J'_1}=\vN\cdot\vtheta,\quad
\theta'_2={\p S\over\p J'_2}=\theta_2,\quad
\theta'_3=\theta_3.
\]
 Since $\theta'_1$ does not evolve in time, the star explores only a
two-dimensional set of its three-dimensional torus. While a star on a
resonant torus does not come arbitrarily close to every point on its torus,
the bigger the numbers $N_j$ are, the more effectively it samples its torus,
and the less likely it is that the resonance condition will be dynamically
important.

\subsection{Perturbation theory}

The importance of resonances only emerges when we consider the effects of
adding a small perturbing Hamiltonian $h(\vtheta,\vJ)$ to the Hamiltonian
$H_0(\vJ)$ for which we have found angle-action variables $(\vtheta,\vJ)$.
We use these coordinates to study the motion of a particle in the full
Hamiltonian $H=H_0+h$.  Hamilton's equations read
 \[
\dot\vtheta={\p H\over\p\vJ}=\vOmega_0+{\p h\over\p\vJ},
\qquad
\dot\vJ=-{\p H\over\p\vtheta}=-{\p h\over\p\vtheta},
\]
 where $\vOmega_0=\p H_0/\p\vJ$.  The perturbing Hamiltonian, like any
function on phase space, is a periodic function of the angles, so we can
Fourier expand it: 
 \[
h(\vtheta,\vJ)=\sum_\vn h_\vn(\vJ)\cos(\vn\cdot\vtheta+\psi_\vn).
\]
 With this expansion, the equation of motion of $\vJ$ is
 \[
\dot\vJ=\sum_\vn \vn h_\vn(\vJ)\sin(\vn\cdot\vtheta+\psi_\vn) 
=\sum_\vn \vn h_\vn(\vJ)\sin(\vn\cdot\vOmega\,t+\psi'_\vn) .
\]
 So long as $\vn\cdot\vOmega\ne0$, the time-averaged value of $\dot\vJ$ vanishes
and we expect $\vJ$ simply to oscillate slightly around its unperturbed
value.  But if a resonance condition is nearly satisfied,
$\vN\cdot\vOmega\simeq0$, the argument of one or more of the sines will
change very slowly, and the cumulative change in $\vJ$ can be significant
even for very small $h_\vN$. Thus resonances permit small forces to act in the
same sense for long times, and hence cause qualitative changes in behaviour.
This is one of the fundamental principles of physics.

Mathematically, we use the new angle variables $\vtheta'$ defined by equation
(\ref{eq:restheta}) and their conjugate actions
\[
J_1={\p S\over\p\theta_1}=J'_1N_1,\quad
J_2={\p S\over\p\theta_2}=J'_1N_2+J'_2,\quad
J_3=J'_1N_3+J'_3.
\]
 Next we express $h$ as a Fourier series in the new angle variables and
discard all terms that involve $\theta'_2$ or $\theta'_3$ on the grounds that
they oscillate too rapidly to have a significant impact on the dynamics.
Since our approximated Hamiltonian depends on neither $\theta'_2$ nor
$\theta'_3$, the conjugate actions $J'_2$ and $J'_3$ will be constants of
motion. The only non-trivial equations of motion are now
 \bea\label{eq:slowmotion}
\dot \theta'_1&=&\Omega'_{01}(J'_1)+\sum_n{\p h'_n\over\p
J'_1}\cos(n\theta'_1+\psi'_{(n,0,0)})\cr
\dot J'_1&=&\sum_n nh'_n(J'_1)\sin(n\theta'_1+\psi'_{(n,0,0)}),
\eea
 where 
\[
\Omega'_{01}={\p H_0(\vJ')\over\p J'_1}=\sum_i\Omega_{0i}{\p J_i\over\p J'_1}
=\Omega_{01}N_1+\Omega_{02}N_2+\Omega_{03}N_3=\vN\cdot\vOmega_0
\]
 and we have suppressed the dependence of $\Omega_{01}$ and $h'_n$ on
$J'_2$ and $J'_3$ because the latter are mere constants.
We have reduced the particle's motion in six-dimensional phase space to
motion in the $(\theta'_1,J'_1)$ plane.

We differentiate with respect to time the first of equations
(\ref{eq:slowmotion})
 \[
\ddot\theta'_1={\p\Omega'_{01}\over\p J'_1}\dot J'_1+\sum_n
\biggl(
{\p^2 h'_n\over\p {J'_1}^2}\dot J'_1\cos(n\theta'_1+\psi'_{(n,0,0)}
)-
n\dot\theta'_1{\p h'_n\over\p J'_1}\sin(n\theta'_1+\psi'_{(n,0,0)}
)\biggr).
\]
 We can neglect the sum in this equation because each of its terms is the
product of a derivative of $h'_n$, which is small, and either $\dot J'_1$,
which is of the same order, or $\dot\theta'_1$, which is also small because
$\Omega'_{01}$ vanishes at the resonance. Therefore we can dramatically
simplify the $\theta'_1$ equation of motion to
\[\label{eq:genpend}
\ddot\theta'_1={\p\Omega'_{01}\over\p J'_1}\dot J'_1
={\p\Omega'_{01}\over\p J'_1}\sum_n nh'_n(J'_1)\sin(n\theta'_1+\psi'_{(n,0,0)}
).
\]
 If we approximate $\p\Omega'_{01}/\p J'_1$ and $h'_n$ by their values on
resonance, and retain only the term for $n=1$ in the sum, we are left with the
equation of motion of a pendulum
 \[\label{eq:pendule}
\ddot\theta=-\omega^2\sin\theta,
\]
 where
 \[
\theta\equiv\theta'_1+\psi'_1 \quad\hbox{and}\quad
\omega^2\equiv-{\p\Omega'_{01}\over\p J'_1}h'_1.
\]

 A pendulum can move in two ways: at low energy its motion is
oscillatory at an angular frequency that falls from $\omega$ at the lowest
energies to zero at the critical energy above which the pendulum circulates.
Consequently, equation (\ref{eq:genpend}) predicts that close to the resonance
(``low energy'') $\theta'_1$ will oscillate. This is the regime of
\deffn{resonant trapping} in which the particle \deffn{librates} around the
underlying resonant orbit. At some critical distance from the resonance
(``high energy'') $\theta'_1$ will start to circulate. Near the threshold
energy, the rate at which $\theta'_1$ advances in time will be highly non-uniform, just
as a pendulum that has only just enough energy to get over top dead centre
slows markedly as it does so. As we move further and further from the resonance,
the rate of advance of $\theta'_1$ becomes more and more uniform, and we
gradually recover the unperturbed motion, in which the rate of advance of
$\theta'_1$ is strictly uniform.

\begin{figure}
\includegraphics[width=.8\hsize]{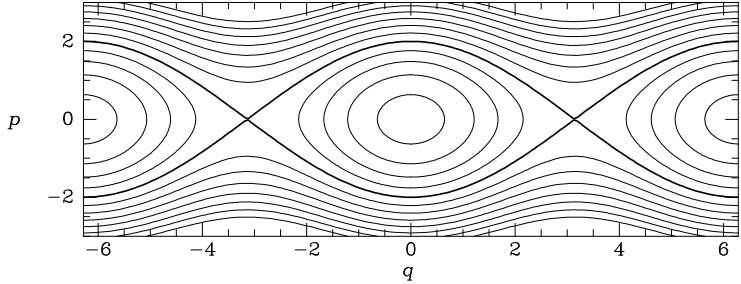}
\caption{The phase plane of a pendulum.  Curves of constant energy $E$
(eq.~\ref{eq:Epend}) are plotted. The particle moves on these, from left to
right in the upper half of the figure and from right to left in the lower
half.}\label{fig:pendule}
\end{figure}

We can obtain a useful pictorial representation of this behaviour by
deriving the energy invariant of equation (\ref{eq:pendule}). We multiply
both sides by $\dot\theta$ to obtain an equation which states that
 \[\label{eq:Epend}
E\equiv\fracj12\dot\theta^2-\omega^2\cos\theta
\]
 is constant. Consequently, the particle moves in the $(\theta,\dot\theta)$
plane along curves of constant $E$. These curves are shown in
\figref{fig:pendule}. The round contours near the centre of the figure show
the motion of a particle that has been trapped by the resonance, while the
contours at the top and bottom of the figure that run from $\theta=-\pi$ to
$\theta=\pi$ show the motion of a particle that continues to circulate.

\begin{figure}
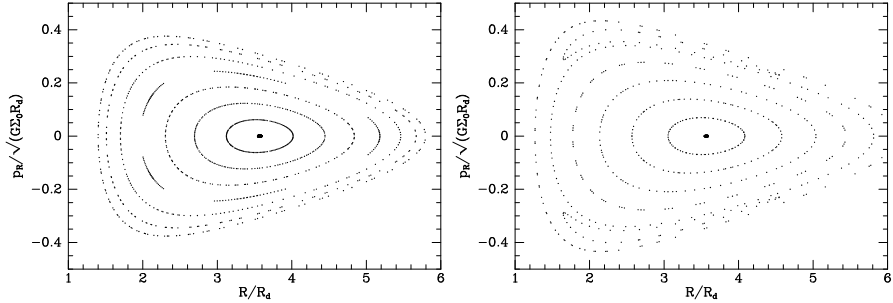

\centerline{\includegraphics[width=.48\hsize]{chapter_binney/fig5a.ps}
\includegraphics[width=.48\hsize]{chapter_binney/fig5b.ps}}
\caption{Surfaces of section for motion in flattened isochrone
potentials: the left panel is for the case of a mass distribution that has
axis ratio $q=0.7$, while the right panel is for $q=0.4$. In the right panel
we see resonant islands generated by the $1:1$ resonance between the radial and
vertical oscillations. No such island is evident in the left panel.}
\label{fig:resonance}
\end{figure}

To appreciate the significance of resonances for galactic dynamics we must
introduce the concept of a \deffn{surface of section}. This is a phase plane
such as the $(R,p_R)$ plane for motion in an axisymmetric potential. Each
time the particle passes through the plane $z=0$ moving upward ($p_z>0$) we
mark the surface of section with a dot at the current values of $R$ and
$p_R$. If the orbit is quasiperiodic, the dots eventually trace out a curve.
This curve is the intersection of the orbit's torus with the $(R,p_R)$ plane,
and in fact the area inside it, $\int \d R\,\d p_R$ is $2\pi$ times the
orbit's radial action $J_r$. \figref{fig:resonance} shows surfaces of section
for motion in two flattened axisymmetric potentials.  In a given panel each
curve is for a different orbit with the same energy and value of $L_z$. Most
of the curves move around a central point. The point itself is made by the
\deffn{shell orbit}, which is two-dimensional and has $J_r=0$; in real space
this orbit is a thin cylindrical shell that has a larger diameter at $z=0$
than at its top or bottom edges. Each curve around this point in
\figref{fig:resonance} is generated by a three-dimensional orbit that forms
an annulus of finite thickness. \figref{fig:one} shows cross sections through
two such annuli.  In \figref{fig:resonance}, the longer an orbit's curve is,
the thicker the annulus and the smaller its vertical extent. The outermost
curve in \figref{fig:resonance} is generated by the orbit $J_z=0$, which is
confined to the plane $z=0$. So in \figref{fig:one} the orbit in the left
panel generates a larger curve in the left panel of \figref{fig:resonance}
than does the orbit of the right panel of \figref{fig:one}.

The right panel in \figref{fig:resonance} is for motion in the potential of a
flatter galaxy than the left panel, and in this panel not all curves loop
around the central point. Two \deffn{resonant islands} have appeared, formed
by orbits that have been trapped by the resonance $\Omega_r=\Omega_z$.

At this point one should be asking ``so what's the perturbation in this case?''
It isn't evident that there is one; all we did was integrate orbits in a
perfectly well defined potential. However, we can imagine that our
Hamiltonian is the sum of a Hamiltonian $H_0$ that would generate a surface
of section in which all curves simply looped around the central point, and a
smaller Hamiltonian $h=H-H_0$, and to ascribe  to $h$ the trapping of orbits
by the $1:1$ resonance. \cite{KaasalainenB94} describe an algorithm that can
be used to generate $H_0$ and $h$.

The resonance just discussed is rather a tame one that probably does not play
a big role in galactic dynamics (but see \S\ref{sec:adiabdef}). Resonances
certainly play a big role in the dynamics of bars. The dynamics of a bar is
strongly affected by the bar's pattern speed, $\Omega_{\rm p}$, the angular
velocity at which the figure of the bar rotates. The bars we see at the
centres of more than half spiral galaxies, including our own, are rapidly
rotating in the sense that in the rotating frame of the bar the dynamics of
most stars depends strongly on the Coriolis force $2\vOmega_{\rm
p}\times\vp$. In cosmological simulations of the clustering of dark matter,
most dark halos are barred and their pattern speeds are so small that the
Coriolis force is unimportant for a significant fraction of their constituent
particles. The gravitational potentials  of galaxies are approximately
isothermal, so let's investigate motion in the potential
 \[\label{eq:logphi}
\Phi=\fracj12v_0^2\ln\left(r_0^2+x^2+{y^2\over q^2}+{z^2\over q_2^2}\right).
\]
 Here $r_0$ is a core radius, within which the potential tends towards that
 of a harmonic oscillator, and $v_0$ would be the circular speed at $x\gg
r_0$ in the $(x,y)$ plane if the axis ratio $q$ were unity. We shall
assume that $q=0.8$, so the potential is slightly elongated along the $x$
axis and for now restrict ourselves to motion in the $(x,y)$ plane so we can
use the device of a surface of section.

\begin{figure}
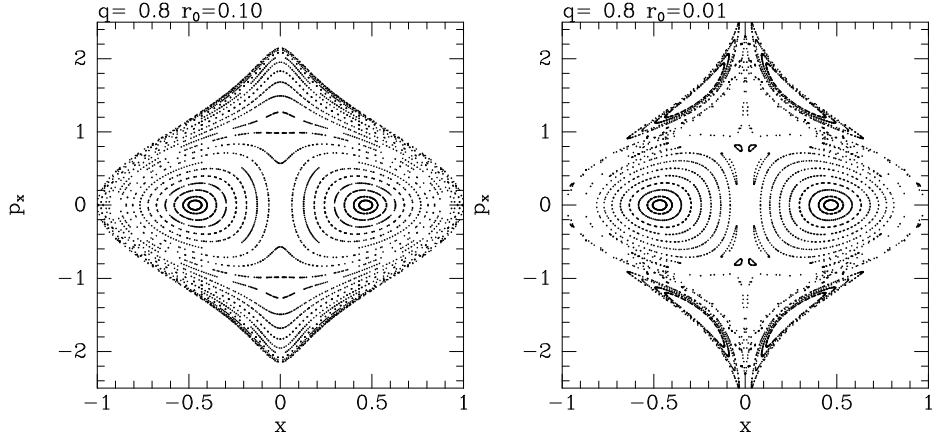

\centerline{\includegraphics[width=.5\hsize]{chapter_binney/fig6a.ps}
\includegraphics[width=.5\hsize]{chapter_binney/fig6b.ps}}
\caption{Surfaces of section for the potential (\ref{eq:logphi}) with axis
ratio $q=0.8$. Left: for $r_0=0.1$; right: for $r_0=0.01$.}\label{fig:logpot} 
\end{figure}

\begin{figure}
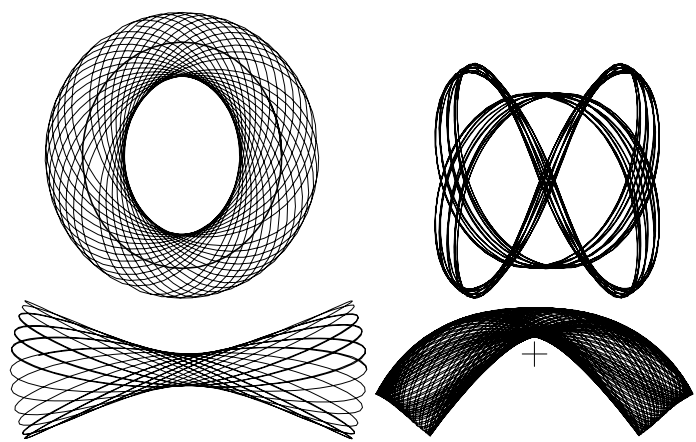

\centerline{\includegraphics[width=.3\hsize]{chapter_binney/fig7a.ps}
\qquad\qquad\includegraphics[width=.25\hsize]{chapter_binney/fig7d.ps}}
\centerline{\includegraphics[width=.39\hsize]{chapter_binney/fig7b.ps}
\includegraphics[width=.35\hsize]{chapter_binney/fig7c.ps}}
\caption{A loop and a box orbit (left); two boxlets
(right). The cross in the lower right panel marks the potential's centre.}\label{fig:loopbox} 
\end{figure}

\figref{fig:logpot} shows $(x,p_x)$ surfaces of section for two values of
$r_0$. In the case $r_0=0.1$ on the left we can identify the curves of two
types of orbit: the two bulls-eyes comprise the curves of \deffn{loop orbits}
such as that depicted in the top left panel of \figref{fig:loopbox}. These
orbits have a well-defined sense of rotation about the $z$ axis. The curves
that surround both bulls-eyes are generated by \deffn{box} orbits such as
that depicted in the bottom left panel of \figref{fig:loopbox}. Box orbits
extend to the potential's centre and do not involve rotation. In the surface
of section on the right of \figref{fig:logpot} we see several resonant
islands in the domain of the box orbits. These islands are made up of orbits
trapped by the resonances $\Omega_y=2\Omega_x$ (outermost island) and
$2\Omega_y=3\Omega_x$ (further in). The right panels of \figref{fig:loopbox}
show the appearance of a couple of these orbits in real space. As this
experiment suggests, the more cuspy a triaxial mass distribution is, the
larger is the fraction of phase that is occupied by resonant box orbits, or
\deffn{boxlets} as they have been called \citep{ValluriM}. Why resonant
trapping becomes more important as the matter distribution becomes more cuspy
is not clear. A fact that is probably relevant is that boxlets generally
avoid the galactic centre, while boxes do not. Given that massive black holes
reside in galaxy centres, the tendency of boxlets to avoid the centre clearly
has important astrophysical consequences by keeping stars safe from being
eaten by the monster there. 

\begin{figure}
\includegraphics[width=.5\hsize]{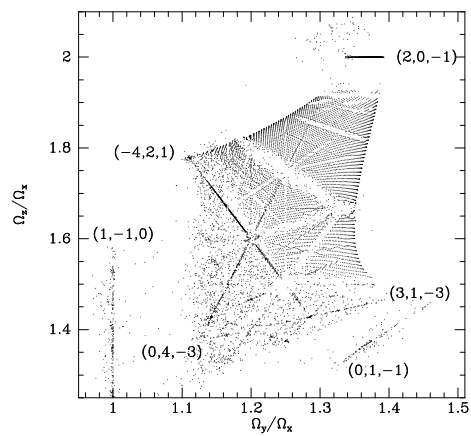}
\caption{Frequency-ratio diagram for the potential (\ref{eq:logphi}) with
$q=0.9$ and $q_2=0.7$.}\label{fig:laskar}
\end{figure}

\subsection{Frequency analysis}

We have seen how surfaces of section give valuable insight into the structure
of phase space. Unfortunately, it is impracticable to construct a surface of
section unless the motion is in two dimensions -- or effectively so in the
case of an axisymmetric potential. Frequency analysis provides a useful way
of determining which resonances play an important role when the potential is
triaxial and the motion of particles is inherently three-dimensional. The
technique consists of numerically integrating orbits and then Fourier
transforming the time dependence of suitable coordinates. From the resulting
spectra one identifies the fundamental frequencies, and determines two
frequency ratios, such a $\Omega_z/\Omega_x$ and $\Omega_z/\Omega_y$. Each
orbit then generates a point in a frequency-ratio diagram such as that of
\figref{fig:laskar}. In this diagram the points of non-resonant orbits fall
along a curvilinear grid, which reflects the systematic way in which the
initial conditions explored phase space. The points of resonant orbits lie
along straight lines. In the vicinity of these lines there is a deficit of
points, because orbits in the underlying integrable potential that would have the
nearly commensurable frequencies that correspond to these locations have been
resonantly trapped so their principal frequencies exactly satisfy a resonance
condition. Actually these orbits still have three independent fundamental
frequencies, but one of these frequencies is the frequency of libration
around the trapping orbit, and this frequency has to be dug out of the
coordinates' spectra and is missed by the simple algorithm from used to identify
the frequencies from which the frequency ratios were calculated.

\begin{figure}
\includegraphics[width=.7\hsize]{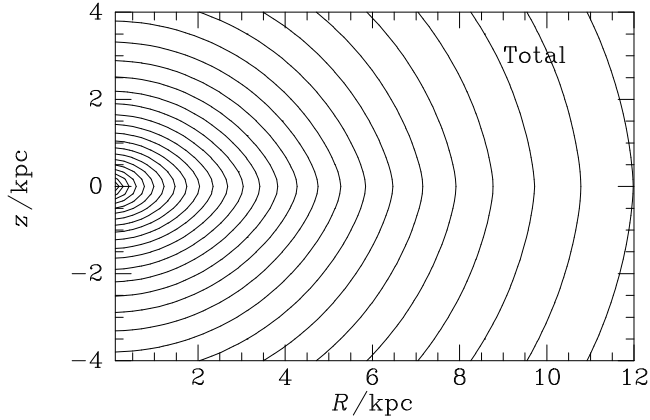}
\caption{A model Galaxy potential. The cuspiness of the equipotential
contours clearly show the extent to which the potential has been flattened by
the mass of the disc.}\label{fig:galpot}
\end{figure}
\subsection{Adiabatic deformation}\label{sec:adiabdef}

The Milky Way's disc has accumulated rather gradually over most of the
Hubble time. As it grew, the Galaxy's potential must have deformed from being
the nearly spherical potential of the dark halo, to a potential that is
significantly pinched towards the plane (\figref{fig:galpot}). Individual
orbits will have distorted in response to the distortion of the potential,
but so long as they remained non-resonant, their actions will have been
invariant: actions are \deffn{adiabtatic invariants}. This fact greatly
facilitates the process of determining the response of a stellar system to
adiabatic distortion of its potential because all we have to do is to move
each star from  its orbit in the original potential to the orbit with the
same actions in the distorted potential. In particular, the structure of the
distorted system depends only on the initial and final configurations, and
not on which configurations it passed through in between.

The story is more complex and interesting if resonant trapping is possible.
When the Galaxy's potential was nearly spherical, every orbit  had a
value of $\Omega_r$ that was bigger than that of either of its other two
frequencies. Stars whose orbits are now confined within a couple of
kiloparsecs of the equatorial plane have $\Omega_r<\Omega_z$, the
frequency associated with motion perpendicular to the plane. So these stars
have at some point satisfied the resonance condition
$\Omega_r=\Omega_z$.  In a flatted potential
$\Omega_r/\Omega_z$ is smallest for orbits that are confined to the
equatorial plane. Therefore the resonance condition was first satisfied by
these orbits.

\figref{fig:resonance} shows surfaces of section for motion in a potential
before and after the resonance condition $\Omega_r=\Omega_z$ is first
satisfied: the islands visible in the right panel are made up of orbits trapped
by this resonance.  Note that the areas of the curves in the left panel are
$2\pi J_r$, so they do not change as the potential flattens. 

The resonance condition is first satisfied by the orbit that is confined to
the equatorial plane; in both panels of \figref{fig:resonance} the curve of
this orbit lies on the outside. Hence the resonant islands first appeared just
inside this curve. As the potential flattened more,
$\Omega_r/\Omega_z$ dropped significantly below unity for the planar
orbit, so the resonance condition was satisfied by orbits with non-zero
$J_z$ and the islands moved inwards. Orbits whose curves lay in the
path of a moving island did one of two things: (a) they were trapped into
the island, or (b) they abruptly increased their radial actions so that their
curves went round the far side of the island.  Which of these two outcomes
happened in an individual case depended on the precise orbital phase of the
star when the potential achieved a particular flattening, but it is most
useful to average over phases and to consider the outcomes to occur with
probabilities $P_{\rm a}$ or $P_{\rm b}$. The  magnitudes of $P_{\rm
a}$ and $P_{\rm b}$ depend of the relative speed with which the island
increased its area and moved: if it simply grew, $P_{\rm a}=1$, and if it
moved without growing $P_{\rm b}=1$. These results follow from Liouville's
theorem that phase-space density cannot increase.

Let's imagine that after a period of stationary growth, the island moved
inwards without growing, and then became stationary while it shrank. In this
case it would have swept up stars with large $J_r$ and small $J_z$ and
released these stars into orbits with smaller $J_r$ and larger $J_z$. In
other words, it will have turned radial motion into vertical motion.
\cite{Touma} have called this process ``levitation''.  Conversely, the moving
island will have reduced the vertical motions and increased the radial
motions of any stars it found in its path through action space. Thus
resonances stirr the contents of phase space. Levitation is a lovely idea but
it's not clear that it is of practical importance. We shall see below that in
a disc analogous scattering by resonances is very important.

\subsection{From order to chaos}

\begin{figure}
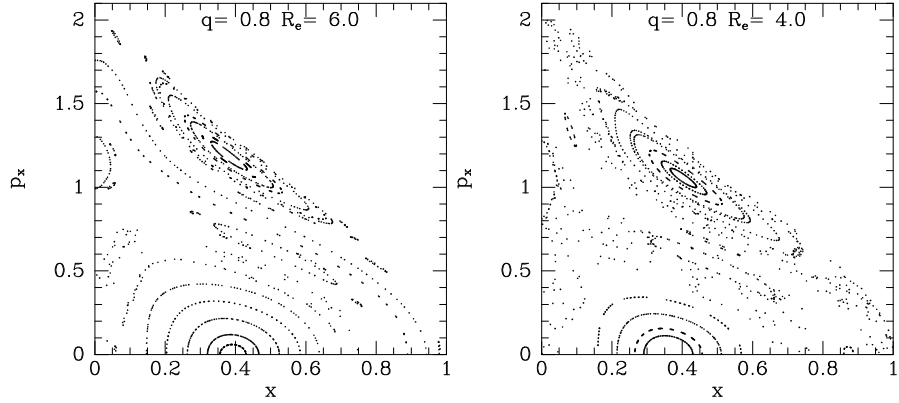

\centerline{\includegraphics[width=.48\hsize]{chapter_binney/fig10b.ps}
\includegraphics[width=.48\hsize]{chapter_binney/fig10a.ps}}
\caption{Surfaces of section for the potential
(\ref{eq:jjbpot}) with $q=0.8$ and $r_0=0.1$. The left panel is for $R_\e=6$
and the right panel is for $R_\e=4$.}\label{fig:jjbpot}
\end{figure}

If we set $z=0$ and express the triaxial potential (\ref{eq:logphi}) in
cylindrical coordinates, it becomes
 \[
\Phi(R,\phi)=\fracj12v_0^2
\ln\left[r_0^2+\fracj12R^2(q^{-2}+1)-\fracj12R^2(q^{-2}-1)\cos2\phi\right].
\]
 Consider now the related potential \citep{Binney82}
 \[\label{eq:jjbpot}
\Psi(R,\phi)=\fracj12v_0^2
\ln\left[r_0^2+\fracj12R^2(q^{-2}+1)-\Big(\fracj12R^2(q^{-2}-1)
+{R^3\over R_\e}\Big)\cos2\phi\right].
\]
 \figref{fig:jjbpot} shows surfaces of section for motion in this potential
when $r_0=0.1$, $q=0.9$ and $R_\e=6$ (left) and $R_\e=4$ (right). Comparing
the left panel of this figure with the left panel of \figref{fig:logpot} we
see that the extra term in the logarithm has greatly increased the number of
resonant islands, and in the right panel we see that it can introduce a
qualitatively new feature: many points seem now to be
scattered at random rather than confined to curves.  Confinement to curves is
an indication that the orbits admit an isolating integral in addition to
energy, and are in consequence quasiperiodic.  Conversely, when the points
are not confined to curves, the orbits lack an additional integral and are
not quasiperiodic. We say these orbits are \deffn{irregular} or
\deffn{chaotic}.

The increase in the number of resonant islands is probably the cause of the
emergence of chaos.  In the surface of section, islands form chains, with the
curves of some non-trapped orbits running on one side of the chain, and those
of others running on the other side of the chain -- \figref{fig:pendule} may
help the reader to picture this situation. The curves of the most nearly
trapped orbits on each side come very close where the islands touch.  The
tiniest perturbation can cause a star on one of these orbits to swap from one
side of the chain to the other. If a star makes such changes, its orbit
ceases to be quasiperiodic. 

If there are several chains of islands in the surface of section, and the
islands of two chains almost touch, a star can make two such swaps, moving
from, say, inside chain 1 to outside chain 3. In this way a star can move
stochastically through a significant region of phase space. This is probably
what happens in \figref{fig:jjbpot}.

\section{Fluctuations}\label{sec:fluctuations}

In previous sections we investigated the orbits of stars in a smooth,
time-independent model of the galaxy's gravitational potential. In reality
the potential contains time-dependent features and in this
section we investigate how these features drive evolution.

A fundamental result is obtained by multiplying the equation of motion
$\dot\vp=-\nabla\Phi$ by $\vp$ and rearranging the result to
 \[
{\d E\over\d t}={\p\Phi\over\p t}.
\]
 Thus stars change their energies if and only if the potential is
time-dependent.  Fluctuations in the potential enable stars to exchange
energy.

\begin{figure}
\includegraphics[width=.5\hsize]{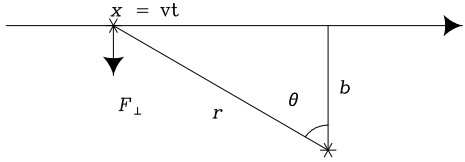}
\caption{A fast encounter between two stars at speed $v$ and impact parameter
$b$. The force perpendicular to the relative velocity is $\sim Gm_1m_2/b^2$
and it acts for a time $\sim 2b/t$.}\label{fig:encounter}
\end{figure}

\subsection{Two-body scattering}

The most obvious source of fluctuations is the moving gravitational
potentials of individual stars. When stars of mass $m_1$ and $m_2$ pass each other at speed $v$
and impact parameter $b$ (\figref{fig:encounter}) the effect is an exchange
of momentum along the line that is perpendicular to the mutual velocity and
has magnitude $\sim2Gm_1m_2/bv$. So the encounter adds to the velocity of
$m_1$ a velocity $\delta\vv_1$ of magnitude $\sim2Gm_2/bv$. The direction of these
increments is random, so we add them in quadrature. The rate of such encounters
is $\sim 2\pi nv b\d b$, where $n$ is the number density of stars, so the
rate of change of $\sum|\delta\vv_1|^2$ is
 \[
{\d\over\d t}\sum|\delta\vv_1|^2=2\pi n\int\d b\,b\left({2Gm_2\over
bv}\right)^2 ={8\pi G^2m_2^2\over v^2}\int{\d b\over b}.
\]
 The integral diverges at both ends of the range of integration. The
divergence at small $b$ is an artifact that can be traced to our use of
$2Gm_2/bv$ as the magnitude of the velocity change in an encounter: an
accurate calculation shows that the velocity change never exceeds $v$
\citep[][eq.~3.53a]{BT08}.
The divergence at large $b$ is real, and indicates that encounters with
impact parameters that are on the order of the size of the system dominate.
Physically, what this means is that the dominant source of fluctuations is
Poisson fluctuations in the number of stars in substantial parts of the
system. The mass inside a volume of radius $r$ will fluctuate by
$\delta M\sim M/\sqrt{N}$, where $N=\fracj43\pi r^3n$ is the number of stars
in this volume. Just outside this volume the gravitational field will
fluctuate by 
\[
\delta g={G\delta M\over r^2}={GM\over r^2\surd N}={Gm_2\over r^2}\surd N=
Gm_2\sqrt{\fracj43\pi{n\over r}}
\]
 This fluctuation acts for a time $\sim r/v$ so it changes the velocity of
any star by
 \[
|\delta\vv|\sim {Gm_2\over v}\sqrt{\fracj43\pi{nr}},
\]
 which grows with $r$, thus confirming that large-scale fluctuations are the
most effective. If we accept that the dominant fluctuations are those
involving half the system, so $r$ is about half the system size $R$, we
conclude that the number of half-crossing times $r/v$ required to change
$\vv$ by of order itself is
 \[
{v^2\over|\delta\vv|^2}\sim {v^4\over
(Gm_2)^2}{1\over{\frac23\pi nR}}.
\]
 The virial theorem implies that $v^2\simeq GM/R$, so
\[
{t_{\rm 2B}\over t_{\rm cross}}
\simeq \left({GM\over
RGm_2}\right)^2{1\over{\frac23\pi nR}} ={(M/m_2)^2\over\frac43\pi
nR^3}=N.
\]
 For a galaxy this number of half-crossing times is many times the age of the
Universe, so the fluctuations associated with the motions of individual stars
are unimportant. But for a globular cluster, which has $N\lta10^5$, $R\sim3\pc$ and
$v\sim6\kms$ so $r/v\sim0.25\Myr$, $\vv$ changes by order itself in
$\sim6\Gyr$ so the process is significant. In an open cluster the process is
even more important.

Two-body interactions randomise the distribution of stars in phase space and
thus drive the system towards thermal equilibrium. No such  equilibrium is
possible for a stellar system that is only confined by its own gravity
\citep[][\S4.10]{BT08}. But
we can understand the impact of two-body interactions by considering the
consequences of trying to reach thermal equilibrium.

The speed $v_\e$ required to escape from a stellar system is never much
larger than the system's characteristic velocity dispersion $\sigma$ -- one
can easily show from the virial theorem that the mass-weighted rms of the
local escape speed is only twice the mass-weighted rms velocity dispersion:
$\langle v_\e^2\rangle^{1/2} =2\langle\sigma^2\rangle^{1/2}$. Consequently,
the velocity distribution is always distinctly non-Gaussian. Two-body
scattering drives the velocity distribution towards Gaussianity, so it is
constantly trying to repopulate the missing tail of the velocity distribution
at $v\gta2\sigma$. Stars scattered into this domain are free and leave the
system, to the system loses mass by \deffn{evaporation} on the two-body
timescale. 

In thermal equilibrium there would be equipartition between the particles. So
massive stars would have a smaller velocity dispersion than low-mass stars.
Consequently, two-body interactions are constantly transferring energy from
more massive to less massive stars, with the consequence that the massive
stars sink towards the centre of the system: two-body scattering drives
\deffn{mass segregation}.

In thermal equilibrium all parts of a body have the same temperature. In a
self-gravitating system there is a tendency for the centre to be hotter than
the outside, if only because the escape speed decreases outwards. So two-body
interactions tend to transfer energy outwards from the core to the envelope.
By the virial theorem, a self-gravitating system that loses energy contracts
and gets hotter, while one that gains energy expands and becomes cooler. So
the conduction of heat from the core to the envelope increases the difference
in temperature between the two parts of the system and accelerates the heat
flow. The upshot is the \deffn{gravithermal catastrophe} in which the core
contracts in both size and mass until it contains only a few stars.

The point to note about evaporation, mass-segregation and the gravithermal
catastrophe is they are all consequences of fluctuations in the gravitational
field driving the system towards an unattainable thermal equilibrium. In star
clusters fluctuations associated with individual stars are sufficient to
generate these effects on astronomically interesting timescales. In galaxies
they are not, but evaporation and the gravithermal catastrophe will be driven
by whatever fluctuations do occur, while equipartition won't be because it
depends on stars of different masses experiencing different fluctuations.
Sources of significant fluctuations in galaxies include giant molecular clouds, spiral
arms, satellite galaxies, and high-speed encounters with other comparable
galaxies.

\subsection{Orbit-averaged Fokker-Planck equation}

In this section we develop a general framework for handling the impact of
fluctuations. The general idea is that, by the strong Jeans theorem
(\S\ref{sec:angleact}) the galaxy's distribution function is at all times a
function $f(\vJ,t)$ of the actions. Fluctuations (and resonances) cause this
function to evolve by causing innumerable small changes $\delta\vJ$ in the
actions of individual stars. Let $P(\vJ,\vDelta)\d^3\vDelta\,\delta t$ be the
probability that in time $\delta t$ a star with actions $\vJ$ is scattered to
the action-space volume $\d^3\vDelta$ centred on $\vJ+\vDelta$. The number of
stars in the action-space volume $\d^3\vJ$ is $(2\pi)^3f(\vJ,t)\d^3\vJ$, so the
number of stars leaving this volume in $\delta t$ is
\[
(2\pi)^3f(\vJ,t)\d^3\vJ\delta t\int\d^3\vDelta\, P(\vJ,\vDelta).
\]
 Similarly, the number of stars that are scattered {\it into\/} this volume
 is
\[
(2\pi)^3\d^3\vJ\delta t\int\d^3\vDelta\, f(\vJ-\vDelta,t)P(\vJ-\vDelta,\vDelta).
\]
 Hence the rate of change of the distribution function is
\[\label{eq:master}
{\p f\over\p
t}=\int\d^3\vDelta\,\big[f(\vJ-\vDelta,t)P(\vJ-\vDelta,\vDelta)-f(\vJ,t)P(\vJ,\vDelta)
\big].
\]
 Since scattering events change actions only slightly,
$P(\vJ,\Delta)$ is appreciable only for $|\vDelta|\ll|\vJ|$. So we can
truncate after just a few terms the Taylor series expansion in $\vJ$ of the product
$f(\vJ,t)P(\vJ,\vDelta)$:
 \begin{eqnarray}
f(\vJ-\vDelta,t)P(\vJ-\vDelta,\vDelta)&=&f(\vJ,t)P(\vJ,\vDelta)\cr
&&\quad-\Delta_i{\p(fP)\over\p J_i}+\fracj12\Delta_i\Delta_j{\p^2(fP)\over\p J_i\p
J_j}+\cdots
\end{eqnarray}
 Substituting the first three terms on the right side of  this expression
into equation (\ref{eq:master}) and cancelling terms, we obtain
 \[\label{eq:FP}
{\p f\over\p t}=-{\p F_i\over\p J_i},
\quad\hbox{where}\quad
F_i\equiv f\overline{\Delta_i}-\fracj12{\p(f\overline{\Delta^2_{ij}})\over\p
J_j},
\]
 \[\label{eq:diffcoeffs}
\overline{\Delta_i}(\vJ)\equiv\int\d^3\vDelta\,\Delta_i P(\vJ,\vDelta)
\quad\hbox{and}\quad
\overline{\Delta^2_{ij}}(\vJ)\equiv\int\d^3\vDelta\,\Delta_i\Delta_j
P(\vJ,\vDelta).
\]
 Equation (\ref{eq:FP}) is the \deffn{orbit-averaged Fokker-Planck equation}.
It states that the rate of change of the distribution function is minus the
divergence of the flux $\vF$ of stars in action space, and we have an
expression for that flux in
terms of the \deffn{diffusion coefficients} defined by equations
(\ref{eq:diffcoeffs}). The latter are simply the expectation value and the
variance of the probability distribution of changes in actions per unit time.

The diffusion coefficients reflect the physics of whatever is responsible for
causing the fluctuations. In some circumstances, for example in a star
cluster, the fluctuations will be
approximately thermal in nature, with temperature $T$. Then the
principle of detailed balance requires that the stellar flux vanish
when the objects
being scattered are in thermal equilibrium with the fluctuations. That is,
$\vF=0$ for
 \[
f(\vJ)=\const\times\e^{-H/kT},
\]
 where $H(\vJ)$ is the Hamiltonian. In this case we have
 \[
{\p f\over\p J_i}=-f{\Omega_i\over kT},
\]
 so $\vF=0$ implies that
\[
0=f\left[\overline{\Delta_i}+\fracj12{\Omega_j\over
kT}\overline{\Delta^2_{ij}}
-\fracj12{\p\overline{\Delta^2_{ij}}\over\p
J_j}\right].
\]
 Clearly the square bracket must vanish, so we obtain an expression for the
first-order diffusion coefficient in terms of the second-order coefficient
\citep{BinneyL}
 \[\label{eq:BinneyLforT}
\overline{\Delta_i}=\fracj12\left({\p\overline{\Delta^2_{ij}}\over\p
J_j}-{\Omega_j\over
kT}\overline{\Delta^2_{ij}}\right).
\]
 This expression is useful because it enables us to obtain the
\deffn{first-order diffusion coefficients} $\overline{\Delta_i}$ from the
\deffn{second-order diffusion coefficients} $\overline{\Delta^2_{ij}}$, and,
while $\overline{\Delta^2_{ij}}$ can be obtained from first-order
perturbation theory (see below), a direct calculation of
$\overline{\Delta_i}$ requires second-order perturbation theory.

The diffusion coefficients are conveniently calculated by expanding the
potential in angle-action coordinates
 \[\label{eq:Phin}
\Phi(\vx,t)=\Phi_0(\vx)+\Phi_1(\vx,t)=\Phi_0
+\sum_\vn\Phi_\vn(\vJ,t)\cos(\vn\cdot\vtheta+\psi_\vn).
\]
 where $\Phi_0(\vx)$ is the potential of the underlying Hamiltonian
$H_0(\vJ)$ and $\Phi_1$ is the fluctuating part of the potential.  Hamilton's
equation of motion for $\vJ$ is
 \[\label{eq:dotJ}
\dot\vJ=-{\p H\over\p\vtheta}=-{\p\Phi_1\over\p\vtheta}
=\sum_\vn\vn\Phi_\vn(\vJ,t)\sin(\vn\cdot\vtheta+\psi_\vn).
\]
 To get a random change in $\vJ$, we need to integrate this equation of motion
for a time $T$ that is longer than the auto-correlation time of the
fluctuations. We do this by expanding the variables in powers of
$\Phi_1/\Phi_0$:
 \[
\vJ(t)=\vJ_0+\vDelta_1(t)+\vDelta_2(t)+\cdots\quad\hbox{and}\quad
\vtheta(t)=\vtheta_0+\vOmega_0t+\vtheta_1(t)+\cdots
\]
 $\vDelta_1$ is obtained by integrating equation (\ref{eq:dotJ}) along the
unperturbed orbit
 \begin{eqnarray}
\vDelta_1(T)&=&\sum_\vn\vn\int_0^T\d t\,
\Phi_\vn(\vJ,t)\sin(\vn\cdot\vtheta+\psi_\vn)\cr
&=&\sum_\vn\vn\int_0^T\d t\,
\Phi_\vn(\vJ,t)\sin[\vn\cdot(\vtheta_0+\vOmega t)+\psi_\vn].
\end{eqnarray}
 To obtain the second-order diffusion coefficient we multiply this equation
by itself and average over initial phases $\vtheta_0$. After reordering the
integrals so $\vtheta_0$ is integrated over first, we find that the innermost
integral is
 \[
(2\pi)^{-3}\int\d^3\vtheta_0\,\sin[\vn\cdot(\vtheta_0+\vOmega_0t)+\psi_\vn]
\sin[\vn'\cdot(\vtheta_0+\vOmega_0 t')+\psi_{\vn'}].
\]
 Using $2\sin A\sin B=\cos(A-B)-\cos(A+B)$ and that the integral of any cosine
that depends on $\vtheta_0$ will vanish, we conclude that the innermost
integral vanishes unless $\vn'=\vn$,\footnote{Since we are using cosine
series, we need sum over only half of $\vn$ space so $\vn'$ will never equal
$-\vn$.} when it's equal to
$\fracj12\cos[\vn\cdot\vOmega_0(t-t')]$. Hence
 \[
\langle\Delta_{1i}\Delta_{1j}(T)\rangle=\fracj12\sum_\vn n_in_j\int_0^T\d t\int_0^T\d
t'\,\Phi_\vn(\vJ,t)\Phi_\vn(\vJ,t')\cos[\vn\cdot\vOmega_0(t-t')].
\]
 Next we take the ensemble average over the fluctuations that are represented
by $\Phi_\vn$. We assume that they are a stationary random process so the
autocorrelation of $\Phi_\vn(\vJ,t)$ depends only on the time lag $t-t'$:
 \[
\overline{\Phi_\vn(\vJ,t)\Phi_\vn(\vJ,t')}=c_\vn(\vJ,t-t').
\]
 with this assumption we have
 \begin{eqnarray}
\overline{\Delta_{1i}\Delta_{1j}}(T)&=&\fracj12\sum_\vn n_in_j
\int_0^T\d t\int_0^T\d
t'\,c_\vn(\vJ,t-t')\cos[\vn\cdot\vOmega_0(t-t')]\cr
&=&\fracj14\sum_\vn
n_in_j\int_{-T}^T\d v\,c_\vn(\vJ,v)\cos(\vn\cdot\vOmega_0v)
\int_{|v|}^{2T-|v|}\d u\\
&=&\fracj12\sum_\vn
n_in_j\int_{-T}^T\d v\,c_\vn(\vJ,v)\cos(\vn\cdot\vOmega_0v)(T-|v|),\nonumber
\end{eqnarray}
 where in the second line we have introduced new coordinates $u=t+t'$ and
$v=t-t'$. Given that we want $T$ to be bigger than the autocorrelation time
of the fluctuations, we have that whenever $c_\vn(\vJ,v)$ is non-negligible,
$|v|\ll T$, so term in the integrand that's proportional
to $|v|$ can be neglected, leaving a result that's proportional to $T$. The
diffusion coefficient is the coefficient of proportionality, so 
 \[\label{eq:Deltfromc}
\overline{\Delta^2_{ij}}=\fracj12\sum_\vn
n_in_j\widetilde c_\vn(\vJ,\vn\cdot\vOmega_0),
\] 
 where $\widetilde c_\vn(\vJ,\omega)$ is the power spectrum of the
fluctuations:
 \[
\widetilde c_\vn(\vJ,\omega)\equiv
\int_{-T}^T\d v\,c_\vn(\vJ,v)\cos(\omega v)
=\int_{-T}^T\d v\,\overline{\Phi_\vn(\vJ,t)\Phi_\vn(\vJ,t-v)}\cos(\omega v).
\]

The bottom line of this result is that the ability of a star to diffuse
through phase space hinges on whether the fluctuations contain power at one
of the star's natural frequencies $\vn\cdot\vOmega_0$. In particular, if the
fluctuations are periodic in time, for example because they arise from a
normal mode of the system, they will drive diffusion only of stars that
resonate with them. In practice periodic fluctuations will simply depopulate
narrow regions of phase space: stars for which $\vn\cdot\vOmega_0$ is equal to the
frequency of the fluctuation will be scattered to new actions and then
cease to be resonant because fundamental frequencies are functions of the
actions. \cite{SellwoodK} find evidence for such action-space ``grooves'' in
numerical simulations of stellar discs 
and show that they can generate new spiral features, which in their turn
generate other grooves.

\begin{figure}
\includegraphics[width=.5\hsize,angle=-90]{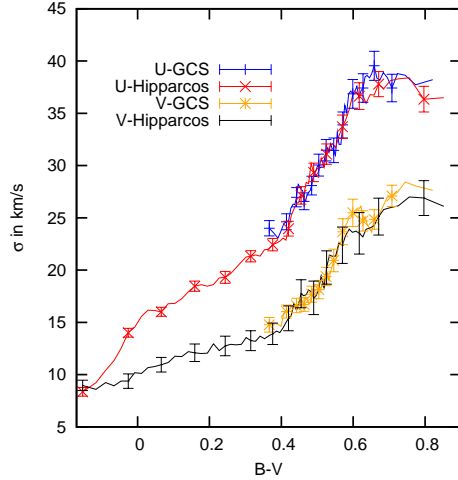}
\caption{The velocity dispersions of Hipparcos stars grouped by colour. From
\cite{AumerB}.}
\label{fig:aumerb}
\end{figure}

\subsection{Heating of the solar neighbourhood}

\figref{fig:aumerb} shows the radial, vertical and azimuthal velocity
dispersions of groups of nearby stars with accurate space velocities as a
function of the colour of the stars.  Blue stars are plotted on the left and
red stars on the right, so all three components of velocity dispersion
increase from blue to red.  Blue stars are massive and short-lived, so in the
blue bins all stars are quite young, while red stars live longer than the age
of the galaxy, so in the red bins we have stars of all ages, but with a bias
to old stars because the star-formation rate was higher in the past than it
is now. So the variation of velocity dispersion with colour indicates that
the random velocities of stars increase over time. From these data and
isochrones one can deduce how velocity dispersion increases with age, and the
conclusion is that $\sigma\sim t^{0.35}$ \citep{AumerB}.

It's instructive to infer from this result how the diffusion coefficients
must scale with $|\vJ|$. We make two simplifying assumptions: (i) that the
dominant scatterers are much more massive than stars, and (ii) that the
velocity dispersions of groups of stars scale with the mean actions in the
group as 
 \[
\sigma_r\propto\sqrt{\langle J_r\rangle}\quad\hbox{and}\quad
\sigma_z\propto\sqrt{\langle J_z\rangle}.
\]
 These relations are exact in the epicycle approximation, in which the radial
and vertical oscillations of stars are harmonic, so for example
$J_r=E_R/\kappa$.\footnote{Quite generally we have that $\Omega_rJ_r$ is equal to
the time-averaged value of $v_R^2$ along any orbit.} Since scattering must be
dominated by giant molecular clouds and spiral arms, the assumption of
massive scatterers will be a good one. In thermal equilibrium with such
massive bodies, stars would have velocity dispersions that are larger than
those of the clouds and arms ($\sim7\kms$) by the square root of the ratio of
masses, so the stars' velocity dispersion would be $>1000\kms$.
Consequently, we can use equation (\ref{eq:BinneyLforT}) in the limit of
infinite temperature,\footnote{See Appendix B of \cite{BinneyL} for a
rigorous justification of this step.} when the Fokker-Planck equation
simplifies to
 \[\label{eq:BinneyL}
{\p f\over\p t}=\fracj12{\p \over\p J_i}\left(\overline{\Delta^2_{ij}}{\p f\over\p
J_j}\right).
\]
 
Stars are born on orbits that have non-negligible angular momenta $L_z\equiv
J_\phi$ but small values of $J_r$ and $J_z$. Consequently, a young population
is initially distributed in action space along the $L_z$ axis, and diffusion
of this population is predominantly away from this line, towards larger
values of $J_r$ and $J_z$. For this reason we neglect derivatives with
respect to $J_\phi$ in equation (\ref{eq:BinneyL}).

In problems involving the ordinary diffusion equation, a key solution is the
Green's function $\exp(-x^2/2t)/(2\pi t)^{1/2}$, which describes the spatial
distribution at time $t$ of particles injected at $x=0$ at time $t=0$.
Analogously, we seek a Green's function of
the form
 \[
f=t^{-2a}f_0(\vX)\quad\hbox{where}\quad
\vX\equiv {\vJ\over t^a}.
\]
 In this solution the mean value of $|\vJ|$ will increase with time as $t^a$,
and the power of $t$ multiplying $f_0$ ensures that the total number of stars
$\int\d L_z\int\d J_r\d J_z\,f$ is conserved as stars diffuse from the axis.  Suppose
$\overline{\Delta^2_{ij}}$ scales such that
$\overline{\Delta^2_{ij}}(k\vJ)=k^b\overline{\Delta^2_{ij}}(\vJ)$. Then
putting $k=t^{-a}$ we have
$\overline{\Delta^2_{ij}}(\vX)=t^{-ab}\overline{\Delta^2_{ij}}(\vJ)$.
Evaluating both sides of equation (\ref{eq:BinneyL}) with these assumptions
yields
 \[
-{1\over t^{2a+1}}\left(2af_0+a\vX\cdot{\p f_0\over\p \vX}\right)=
\fracj12 t^{ab-4a}{\p\over\p X_i}\left(\overline{\Delta^2_{ij}}(\vX){\p f_0\over\p X_j}\right).
\]
 This equation can be valid at all times only if $2a+1=4a-ab$, so $b=2-1/a$.
Consequently, the empirical result $\langle J_r\rangle\sim\sigma_r^2\sim
t^{2/3}$ implies $a\simeq\fracj23$ and $b\simeq\fracj12$. 

The scaling $\sigma_r\sim t^{1/2}$, which has been advocated by \cite{Wielen}
and several subsequent authors, implies $a=b=1$. A simple argument shows that
it is implausible for the diffusion coefficients to grow so rapidly with
$|\vJ|$. In the epicycle approximation, $J_r$ differs from the epicycle
energy $E_R$ only by the (constant) epicycle frequency, so $\Delta_r\sim
\Delta E_R=\vv\cdot\delta\vv$, where $\delta\vv$ is the projection into the
equatorial plane of the change in a star's velocity as a result of a
scattering event. Hence $\langle\Delta_r^2\rangle\sim |\vJ|$ implies
 \[
E_R\sim\langle(\Delta
E_R)^2\rangle\sim\langle (\vv\cdot\delta\vv)^2\rangle\sim\langle E_R|\delta
\vv|^2\rangle.
\] 
 That is, $\sigma_r\sim t^{1/2}$ implies that $|\delta\vv|$ is independent of
$|\vv|$. However, gravitational scattering always causes the momentum change
$\delta\vv$ to decrease with increasing speed because the gravitational force
is independent of speed and the time for which it acts decreases as
$1/|\vv|$. 

Can we derive $\overline{\Delta^2_{ij}}(k\vJ)\sim
k^{1/2}\overline{\Delta^2_{ij}}(\vJ)$ from physics? \cite{BinneyL} show that
this scaling {\it is\/} predicted by the model of cloud-star scattering that
was introduced by \cite{SpitzerS}. However, this model is defective in two
respects: (i) it assumes that the relative velocity with which a star
encounters a cloud is dominated by epicycle motion rather than differential
rotation, and, more seriously, (ii) it assumes that stars are confined to the
equatorial plane. In reality as a star ages it oscillates with increasing
amplitude and period perpendicular to the plane, and these oscillations
decrease its probability of being scattered by a cloud. Consequently, when
this effect is taken into account, $\overline{\Delta^2_{ij}}(\vJ)$ increases
with $|\vJ|$ more slowly than as $|\vJ|^{1/2}$.

\cite{BinneyL} show that three-dimensional scattering by molecular clouds
generates a  tensor of diffusion coefficients $\overline{\Delta^2_{ij}}$
which is highly anisotropic. The consequence of this anisotropy is that we
expect $\sigma_z/\sigma_r\sim0.8$, which is significantly larger than the
observed value, $\sim0.6$.  \cite{Sellwood} argues that the discrepancy
arises from the erroneous assumption of an isotropic distribution of
encounters: as in two-body scattering, distant encounters are important, and
since both stars and clouds lie within the disc, distant encounters are
dominated by the velocity components that lie within the plane and do not
change $J_z$.

Thus it seems that scattering of stars by giant molecular clouds may
set the ratio of the vertical and horizontal velocity dispersions of disc
stars. While star-cloud scattering makes a significant contribution to the secular
increase in the velocity dispersions of stars, it probably cannot account
fully for the data because its effectiveness declines rapidly with
increasing velocity dispersion and thus cannot account for the
numbers of stars with radial dispersions $\gta30\kms$.

\section{Spiral structure}

Thin galactic discs are extremely prone to generating spiral structure. Many
manifestations of spirality arise from gas-dynamics rather than stellar
dynamics -- for example the chains of blue stars so evident in blue and UV
exposures, and the spiral distributions of \hi\ and CO detected at radio
frequencies. In such cases the old stellar disc is believed to carry a
large-amplitude spiral also, but it is not easy to detect these a stellar
spirals. It is most evident in near IR photometry, which is dominated by a
combination of stars near the top of the giant branch and low-mass
main-sequence stars \citep{RixZ}. The former are a measure of recent star
formation so they tell us more about gas than stellar dynamics, but the
latter contain most of the mass of the disc, so are a window into the key
dynamics. \cite{RixZ} find that the $2.2\,\mu$m surface brightnesses of spiral
galaxies carry spiral structures that have amplitudes of order unity, and that
the fluctuation in the surface density of stars is probably nearly as large.
Radio-frequency spectral lines and the H$\alpha$ line show that spiral
disturbances are associated with streaming velocities $\sim7\kms$.

Spiral structure proves to be an intrinsically non-linear phenomenon, and as
a consequence our understanding of it is still frustratingly incomplete. It's
raison d'\^etre is, however, clear: it is the principal means by which
galaxies transport angular momentum outwards, which enables them to increase
their entropy -- i.e., heat their discs. We start our study of spiral
structure by examining how this heating comes about.

\begin{figure}
\includegraphics[width=.6\hsize]{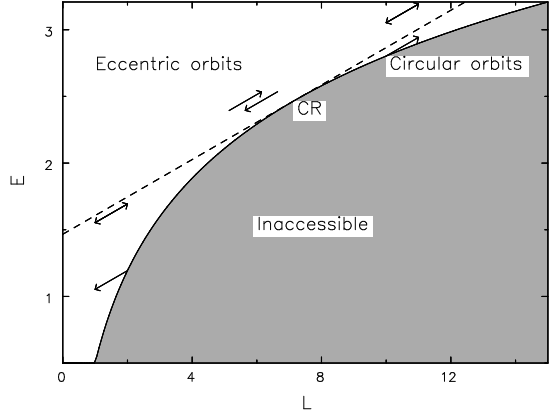} \caption{Energy
versus angular momentum for planar orbits in an axisymmetric potential -- a
``Lindblad diagram''. No orbits lie in the shaded area, which is bounded by
the points of circular orbits. A potential that is stationary in a rotating
frame moves stars along lines with slope $\d E/\d L_z=\Omegap$. (From
Sellwood \& Binney 2002)}
\label{fig:Lindblad}
\end{figure}
\subsection{Secular evolution driven by spiral structure}

Let's assume that spiral structure is a nearly stationary pattern that
rotates at some fixed angular speed $\Omegap$. In this case, when we write
the potential of the spiral structure in angle-action variables
(eq.~\ref{eq:Phin}), the expansion coefficients $\Phi_\vn(\vJ,t)$ will
contain only multiples of $\Omegap$ in their temporal Fourier transforms.
Hence the power spectrum of the potential $\widetilde c_\vn(\vJ,\omega)$,
which appears in equation (\ref{eq:Deltfromc}) for the diffusion
coefficients, will be non-zero only when $\omega$ is equal to one of these
frequencies, so the spiral will enable only a minority of stars to diffuse.
We now examine the impact of the spiral on the minority stars that resonate
with it.

If we work in the frame of reference that rotates at frequency $\Omegap$, the
motion of each star is governed by a time-independent Hamiltonian, the
numerical value of which, the \deffn{Jacobi constant}, is an isolating
integral. In terms of the energy $E$ of motion in the non-rotating frame, the
rotating-frame Hamiltonian is
 \[
H=E-\Omegap L_z.
\]
 Since $H$ is an integral, $\d H=0$ and changes in $E$ and $L_z$ caused by the spiral satisfy
 \[\label{eq:Lindblad}
\d E=\Omegap\,\d L_z.
\]
 \figref{fig:Lindblad} is a plot of $E$ versus $L_z$, and equation
(\ref{eq:Lindblad}) states that in this figure a steadily rotating spiral
moves stars on lines of slope $\Omegap$. The physically accessible region is
bounded below by the locus of circular orbits, which are the orbits with the
largest value of $L_z$ for each given $E$, so there are no orbits in the
shaded region below this boundary. The slope of the boundary, $(\p E/\p
L_z)_{J_r=0}$, is the circular frequency $\Omega(L_z)$. Clearly at the
\deffn{corotation resonance} (CR), where $\Omega(L_z)=\Omegap$, the spiral
scatters stars from one circular orbit to another. Elsewhere, the spiral
scatters stars away from the boundary, to places where the energy exceeds
that of the circular orbit of the given value of $L_z$, and the additional
energy will be invested in epicyclic motion. Inside the CR, the angular
momenta of stars must be reduced, while outside the CR it must be increased.
Thus the spiral must move angular momentum outwards.

We have seen that significant shifts in actions only occur at
resonances, where $\vn\cdot\vOmega_0=m\Omegap$, where $m$ is the number of arms
that the spiral has because the time dependence of the potential is
$\propto\cos(m\Omegap t+\psi)$. Besides the CR [$\vn=(0,m,0)$], the two most
important resonances are the \deffn{inner Lindblad resonance} (ILR), where
$\vn=(-1,m,0)$, and the \deffn{outer Lindblad resonance} (OLR), where
$\vn=(1,m,0)$. At the ILR the D\"oppler-shifted frequency at which a star
perceives the spiral is $m(\Omega-\Omegap)$ and this coincides with its radial frequency,
$\Omega_r$, while at the OLR the perceived frequency of the spiral is
$m(\Omegap-\Omega)$ and this again coincides with $\Omega_r$. We have shown
that the  spiral absorbs $L_z$ at the ILR and emits it at the OLR. At both
places changes in $L_z$ heat the disc.

At the CR the change in angular momentum can have either sign, but the
star simply moves from one circular orbit to another, so the disc is not
heated. In fact these shifts of stars are so inconspicuous that for decades
they were overlooked. They are astronomically important, however, because in
galactic discs metal abundances generally decrease outwards, so the radial
migration of stars at the CR can be detected if metallicities are measured.
Specifically, radial migration ensures that at each radius there are stars of
the same age but differing metal abundances because they formed at different
radii.

\begin{figure}
\centerline{\includegraphics[width=.4\hsize]{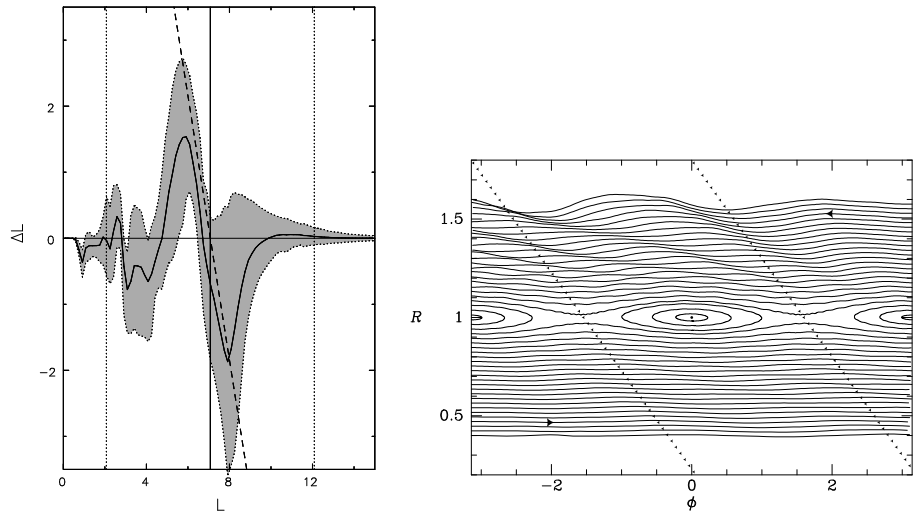}
\quad\includegraphics[width=.55\hsize]{chapter_binney/fig14b.ps}}
\caption{Left: the distribution of changes in angular momentum amongst the
first 20\% of stars in an N-body simulation when they are ordered by initial
epicycle energy. The vertical lines show the angular momenta corresponding to
the ILR, CR and OLR of the spiral pattern within the simulation. The full
curve shows the mean of the distribution and the shaded region is bounded by
its 20th and 80th percentiles. The dashed line has a slope of $-2$.
The right panel shows the response of initially circular orbits to a
transient spiral potential (after Sellwood \& Binney 2002).}\label{fig:SellB}
\end{figure}

\cite{SellwoodB} explored these phenomena with N-body simulations of discs.
In one experiment they introduced an action-space groove into the initial
conditions to generate an isolated, transient spiral feature. The left panel of
\figref{fig:SellB} shows the distribution of ensuing changes in $L_z$ versus
initial $L_z$. Vertical lines mark the locations of the CR and Lindblad
resonances for the measured pattern speed. Stars interior to the CR
gain $L_z$ and transfer to outside the CR, while those outside the CR lose $L_z$ and
move inwards. Thus these stars swap places. The right panel explains how this
is done by plotting orbits in the $(\phi,R)$ plane when a steady two-armed
perturbation is imposed. There are islands formed by orbits that are trapped
at the resonance, and wavy lines of orbits that continue to circulate. Orbits
forming the island constantly move from inside to outside the CR and back again.
When a spiral potential emerges, it creates islands which grow with the
potential by sweeping up orbits from the wavy regions. When the potential
fades, the islands shrink and the trapped orbits are released on each side. A
star that was on an inner wavy orbit at entrapment may  move on a trapped
orbit from inside the
CR to outside the CR before being released to a wavy orbit
outside the CR. The maximum distance stars can move their guiding centres is set
by the largest extent of the islands.  Stars that start far inside the CR are
released far outside the CR. The tendency for the populated regions in the left
panel of \figref{fig:SellB} to slope from top left to bottom right with
gradient $-2$ confirms this picture.  \cite{SellwoodB} called this process of
swapping places around CR \deffn{churning}.

\begin{figure}
\includegraphics[width=.65\hsize]{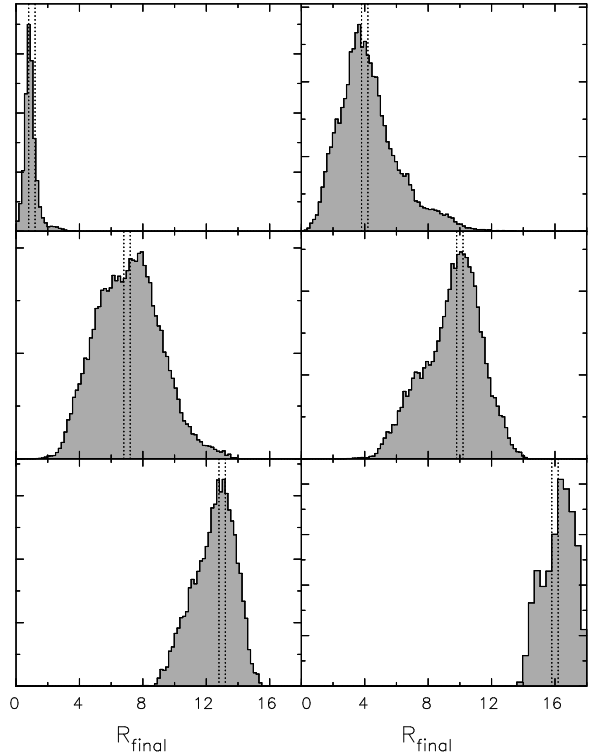} \caption{Each
panel shows the distribution of the final guiding-centre radii of the stars
in a disc simulation whose initial guiding-centre radii lay within the region
between the dotted vertical lines. The disc had a flat rotation curve,
$Q=1.5$, and half of the radial force was provided by a fixed halo. The
duration of the simulation was $\sim4\Gyr$. (From
Sellwood \& Binney 2002).}\label{fig:beforeafter}
\end{figure}

In another experiment Sellwood \& Binney allowed a disc to evolve for
$\sim5\Gyr$ without seeding any spiral structure. Irregular and quite weak
spiral structure emerged as the disc was evolved.  \figref{fig:beforeafter}
shows histograms of the final radii of stars that all started from the narrow
radial bands marked. Stars that started at the current radius of the Sun
finished at radii that are often $1.2-2\kpc$ from $R_0$. From the strength of
spiral structure seen in NIR photometry, Sellwood \& Binney estimated that
over the Hubble time stars will typically migrate $\sim2\kpc$ from their
birth radii.

\subsection{Propagation of spiral waves}\label{sec:tightwind}

If we disturb the surface of a pond with a stone, the water molecules hit by
the stone disturb water molecules slightly further away, which in their turn
disturb their neighbours, and in this way much of the energy of the original
impact is carried over the surface of the pond to be dissipated as the waves
hit the pond's shore. If we disturb the stars in some region of a disc by
perturbing the gravitational potential in their neighbourhood, the orbits of
these stars will change, and a short time later their contribution to the
disc's density distribution will change, which will modify the disc's
potential. This modified potential will disturb the orbits of other stars,
and in this way a way of disturbance will propagate over the disc.

A significant difference between the pond and the disc is that in the pond
water molecules excite their neighbours through pressure: they simply push on
molecules they touch; the interaction is {\it local}. In the disc stars
disturb other stars by modifying the gravitational field, which propagates at
the speed of light. The latter is effectively infinite: we are dealing with action
at a distance, so the physics is inherently {\it non-local}. If we take this
non-locality seriously, we have to compute globally and proceed straight to
the determination of the normal modes of the entire disc.

There is, however, a special case in which the gravitational interaction is
effectively local. This is the case of \deffn{tightly wound} waves: as one
moves away from a density wave in the disc, the latter's gravitational field
decays on the lengthscale of a wavelength because, by virtue of the
long-range nature of gravity, the positive and negative contributions to the
field at the point of observation from peaks and troughs soon cancel rather
precisely. If waves in a disc have a wavelength that is small compared to the
local radius, their gravitational field is highly localised.  For such waves
we can derive a dispersion relation and thus obtain the group velocity, etc.

We now set up the equations whose solution yields the normal modes of a
stellar disc. Then instead of solving them we use the tight-winding
approximation to derive the dispersion relation of tightly wound  spiral
waves. Finally we use this dispersion relation to gain insight into the
global dynamics of self-gravitating discs.

We start by finding how the DF $f(\vJ)$ is changed by a perturbing potential
$\delta\Phi(\vx)$. The governing equation is the collisionless Boltzmann
equation
 \[
{\p f\over\p t}=[H,f]\equiv{\p H\over\p\vtheta}\cdot{\p f\over\p\vJ}
-{\p H\over\p\vJ}\cdot{\p f\over\p\vtheta},
\]
 where $[..]$ denotes the Poisson bracket. Writing
$f(\vJ,\vtheta,t)=f_0(\vJ)+ f_1(\vJ,\vtheta,t)$ and
$H(\vJ,\vtheta,t)=H_0(\vJ)+H_1(\vJ,\vtheta,t)$, we have
to first order
 \[\label{eq:focbe}
{\p f_1\over\p t}=[H_0,f_1]+[H_1,f_0]=
 -{\p H_0\over\p\vJ}\cdot{\p f_1\over\p\vtheta}+
{\p H_1\over\p\vtheta}\cdot{\p f_0\over\p\vJ}.
\]
 We note that $\p H_0/\p\vtheta=\vOmega_0$ and Fourier expand $H_1$ and
$f_1$
 \begin{eqnarray}
 f_1(\vJ,\vtheta,t)&=&\sum_\vn\delta f_\vn(\vJ)\e^{\i(\vn\cdot\vtheta-\omega
t)}\cr
H_1(\vJ,\vtheta,t)&=&\sum_\vn\delta\Phi_\vn(\vJ)\e^{\i(\vn\cdot\vtheta-\omega
t)},
\end{eqnarray}
 where by virtue of the time-translation invariance of
equation (\ref{eq:focbe}) we have assumed that all perturbed quantities have
the same frequency, $\omega$.  Using these expansions in equation
(\ref{eq:focbe}) we equate coefficients of $\e^{\i(\vn\cdot\vtheta-\omega
t)}$ on each side to obtain
 \[\label{eq:fnfromPhin}
\delta f_\vn={\vn\cdot\p f_0/\p\vJ\over\vn\cdot\vOmega-\omega}\delta\Phi_\vn.
\]
 
The frequencies $\omega$ of the system's normal modes are determined by the
requirement that $\Phi_1$ is related by Poisson's equation to the density
fluctuation implied by $f_1$. This requirement reads
 \[\label{eq:normalm}
{1\over4\pi
G}\sum_\vn\nabla^2\left(\delta\Phi_\vn\e^{\i\vn\cdot\vtheta}\right)
=\e^{\i\omega t}\int\d^3\vp\,f_1=\sum_\vn\int\d^3\vp\,{\vn\cdot\p
f_0/\p\vJ\over\vn\cdot\vOmega-\omega}\delta\Phi_\vn\e^{\i\vn\cdot\vtheta}.
\]
 This equation is homogeneous in $\delta\Phi_\vn$ so we expect it to have
non-trivial solutions only for particular values of $\omega$, the frequencies
of normal modes.  Finding the normal-mode frequencies is hard because the
equation involves in an essential way two systems of phase-space coordinates:
$(\vx,\vp)$ and $(\vtheta,\vJ)$.

Although equation (\ref{eq:normalm}) can be tackled \citep[see for
example][]{Kalnajs,ReadE}, we will pursue a simpler course. We restrict ourselves to
razor-thin discs, which have a four-dimensional phase space, and use
the tight-winding approximation. We
assume that the disturbance is a tightly wound spiral wave, so
 \[
\Phi_1(R,\phi,t)=\epsilon\e^{\i(kR+m\phi-\omega t)}
\]
 with $kR\gg1$ for trailing waves and $kR\ll-1$ for leading waves. It is
straightforward to show from Poisson's equation that the corresponding
surface density is
 \[\label{eq:SigfromPoiss}
\Sigma_1=-{|k|\over2\pi G}\Phi_1.
\]
 We adopt the epicycle approximation (\S\ref{sec:epicycle}), in which the
real-space coordinates are related to angle-action coordinates by
 \[\label{eq:epicycJtheta}
R=\Rg+a\cos\theta_r,\quad \phi=\theta_\phi+{\gamma a\over\Rg}\sin\theta_r,
\]
 where
\[
a\equiv\sqrt{2J_r\over\kappa}\quad\hbox{and}\quad \gamma\equiv2\Omega/\kappa.
\]
 Then the Fourier decomposition of
$\Phi_1$ is
 \begin{eqnarray}\label{eq:needBessel}
\Phi_1(R,\phi,t)&=&\epsilon\e^{\i(kR+m\phi-\omega t)}\cr
&=&\epsilon\e^{\i k\Rg}\exp(\i ka\cos\theta_r)\e^{\i
m\theta_\phi}\exp\left(\i{m\gamma a\over\Rg}\sin\theta_r\right)\e^{-\i\omega
t}.
\end{eqnarray}
 We now combine the two exponentials of circular functions into a single
exponential of a single cosine and use equation (8.511.4) of \cite{GradstynR}
to express this as a sum over Bessel functions $\cJ_l$.
Specifically\goodbreak
 \begin{eqnarray}\label{eq:startBessel}
\exp\left[\i\left(ka\cos\theta_r+{m\gamma
a\over\Rg}\sin\theta_r\right)\right]&=&\exp\left[\i
a\cK\sin(\theta_r+\alpha)\right]\cr
&=&\sum_l\cJ_l(\cK a)\e^{\i l(\theta_r+\alpha)}
\end{eqnarray}
where  $\alpha$ is the pitch angle and $\cK$ is the total wavenumber of the
spirals:
\[\label{eq:defsK}
\alpha(J_\phi)\equiv\arctan\left({m\gamma\over k\Rg}\right)
\quad\hbox{and}\quad
\cK(J_\phi)\equiv\sqrt{k^2+{m^2\gamma^2\over\Rg^2}}.
\]
 Using equation (\ref{eq:startBessel}) to rewrite equation
(\ref{eq:needBessel}) we obtain
\[\label{eq:lesH}
\Phi_1(\vtheta,\vJ,t)=\epsilon\sum_{l=-\infty}^\infty\e^{\i(k\Rg+l\alpha)}
\cJ_l(\cK a)\e^{\i(l\theta_r+m\theta_\phi-\omega t)},
\]
  Equation (\ref{eq:lesH}) tells us what $\delta\Phi_\vn(\vJ)$ is for
$\vn=(l,m,0)$:
 \[
\delta\Phi_{(l,m,0)}=\epsilon\e^{\i(k\Rg+l\alpha)}\cJ_l(\cK a).
\]
 Using this in equation (\ref{eq:fnfromPhin}) we obtain the change in the DF
caused by $\Phi_1$: 
\[\label{eq:fndisc}
\delta f_\vn={\vn\cdot\p f_0/\p\vJ\over\vn\cdot\vOmega-\omega}
\epsilon\e^{\i(k\Rg+l\alpha)}\cJ_l(\cK a)\qquad[\vn=(l,m,0)].
\]

The change in the surface density is $\Sigma_1=\int\d^2\vp f_1$. Since
our expression for $f_1$ uses angle-action coordinates rather than
$(\vx,\vp)$ coordinates, we use a trick based on the fact that
$\d^2x\d^2\vp=\d^2\vtheta\d^2\vJ$ because both systems are canonical:
 \begin{eqnarray}
\Sigma_1(R,\phi,t)&=&\int\d^2\vp f_1={1\over R}\int\d^2\vp\int\d
R'\,R'\int\d\phi'\,\delta(\phi-\phi')\delta(R-R')f_1\cr
&=&{1\over R}\int\d^2\vJ\d^2\vtheta \,\delta(\phi-\phi')\delta(R-R')f_1\\
&=&{1\over R}\int\d^2\vJ\d^2\vtheta
\,\delta(\phi-\phi')\delta(R-R')\sum_\vn\delta 
f_\vn\e^{\i(\vn\cdot\vtheta-\omega t)}.
\nonumber
\end{eqnarray}
 Using equation (\ref{eq:epicycJtheta}) for $\phi'$ and $R'$, and equation
(\ref{eq:fndisc}) for $\delta f_\vn$, this becomes
 \begin{eqnarray}
\Sigma_1(R,\phi,t)&=&{\epsilon\over R}
\int\d^2\vJ\d^2\vtheta
\,\delta\Bigl(\phi-\theta_\phi-{\gamma a\over\Rg}\sin\theta_r\Bigr)
\delta(R-\Rg-a\cos\theta_r)\cr
&&\times\sum_{\vn=(l,m,0)}{\vn\cdot\p f_0/\p\vJ\over\vn\cdot\vOmega-\omega}
\e^{\i(k\Rg+l\alpha)}\cJ_l(\cK a)\e^{\i(\vn\cdot\vtheta-\omega t)}.
 \end{eqnarray}
 The two Dirac delta-functions enable us to carry out the integrals over
$\theta_\phi$ and $J_\phi$. This done every occurrence of $\Rg(J_\phi)$
(including those in the definitions of $\kappa$, $a$, etc.) should, strictly,
be replaced by $R-a\cos\theta_r$. However, the tight-winding approximation
allows us to neglect the small difference between $\Rg$ and $R$ except when
it occurs in the argument of an exponential multiplied by the large number
$k$. With the aid of this approximation we obtain
 \begin{eqnarray}\label{eq:dSigma}
\Sigma_1(R,\phi,t)&\simeq&{\epsilon\over R}\e^{\i(kR+m\phi-\omega t)}{\d
J_\phi\over\d\Rg}\bigg|_{\Rg=R}
\sum_{l=-\infty}^\infty\e^{\i l\alpha}\int\d J_r
\,\cJ_l(\cK a){\vn\cdot\p f_0/\p\vJ\over\vn\cdot\vOmega-\omega}\cr
\times\int\d\vtheta_r&&\hskip-10pt\exp\left[\i\left(l\theta_r-m{\gamma a\over
R}\sin\theta_r-ka\cos\theta_r\right)\right]
\quad[\vn=(l,m,0)].
\end{eqnarray}
 It is simple to show that $\d J_\phi/\d\Rg\equiv\d
 L_z/\d\Rg=\Rg\kappa/\gamma$. For $f_0$ we adopt 
\[
f_0(\vJ)={\gamma\Sigma_0\over2\pi\sigma^2}\,\e^{-\kappa J_r/\sigma^2},
\]
 which with the epicycle approximation (eq.~\ref{eq:epicycJtheta}) yields the
Schwarzschild velocity distribution with radial dispersion $\sigma$
\citep[e.g.~\S4.4.3][]{BT08,Binney10}. Finally, we
use  equation (\ref{eq:startBessel}) to express the exponential of
sinusoids in the last line of equation (\ref{eq:dSigma}) as a sum over Bessel
functions. Then we can evaluate the integral over $\theta_r$ to obtain
 \begin{eqnarray}\label{eq:SigmaPhi}
\Sigma_1(R,\phi,t)&=&{\epsilon\kappa^2\Sigma_0\over\sigma^4}\,
\e^{\i(kR+m\phi-\omega t)}\sum_{l=-\infty}^\infty{-l\over
l\kappa+m\Omega-\omega}\int\d J_r\,|\cJ_l(\cK a)|^2\e^{-\kappa J_r/\sigma^2}\cr
&=&{\epsilon\kappa\Sigma_0\over\sigma^2}\,
\e^{\i(kR+m\phi-\omega t)}\sum_{l=-\infty}^\infty{-l I_l(\chi)\e^{-\chi}\over
l\kappa+m\Omega-\omega},
\end{eqnarray}
 where equation (6.615) of \cite{GradstynR} has been used to evaluate the
integral over $J_r$, $I_l$ is a modified Bessel function, and
\[
\chi\equiv{\cK^2\sigma^2\over\kappa^2}.
\]
 Two more definitions and the identity $I_l(z)=I_{-l}(z)$ enable us to write
equation (\ref{eq:SigmaPhi}) in the neater form
 \begin{eqnarray}\label{eq:SigmaPhin}
\Sigma_1(R,\phi,t)&=&{\cK^2\Sigma_0\over\kappa^2(1-s^2)}\cF(s,\chi)\Phi_1,
\end{eqnarray}
 where
 \[
s\equiv{\omega-m\Omega\over\kappa},\quad
\cF(s,\chi)\equiv2(1-s^2){\e^{-\chi}\over\chi}\sum_{l=1}^\infty{I_l(\chi)\over1-s^2/l^2}.
\]
 The final step is to require that the value of $\Sigma_1$
from equation (\ref{eq:SigmaPhin}) agrees with that given by equation
(\ref{eq:SigfromPoiss}). Eliminating $\Sigma_1$ between these equations
and approximating $\cK$ by $|k|$ (see eq.~\ref{eq:defsK}), we obtain the
Lin-Shu-Kalnajs dispersion relation for tightly wound spiral waves:
 \[\label{eq:LSK}
{|k|\over k_{\rm
crit}}\cF(s,\chi)=(1-s^2)=1-{(\omega-m\Omega)^2\over\kappa^2},
\quad\hbox{where}\quad
k_{\rm crit}\equiv{\kappa^2\over2\pi G\Sigma_0}.
\]

 For an $m$-armed spiral with pattern speed $\Omegap$, $\omega=m\Omegap$ so 
 \[
s={m(\Omegap-\Omega)\over\kappa},
\]
 which rises from $-1$ at the ILR through zero at the CR to $1$ at the OLR. Hence
from equation (\ref{eq:LSK}) $k\cF$ vanishes at the Lindblad resonances. One
finds that $\cF$ behaves roughly as $k(1-b k)$ with $b>0$, so the left side
of the dispersion relation peaks for some $k$, and \cite{Toomre64} showed
that if the disc is stable to axisymmetric disturbances this peak value is
smaller than unity, so solutions for $k$ cannot be found for a range of small
values of $s^2$. Thus waves are forbidden in a zone around the CR as well as
inside the ILR and outside the CR. In the permitted zones {\it two} values of
$k$ can be found for given $s$, the values approaching one another as $s$
approaches the forbidden zone around the CR. Thus there are two
branches to the dispersion relation, and the branches merge at the edge of
the CR zone.

When \cite{Toomre69} determined the group velocity of waves from the
dispersion relation, he found that short-leading waves propagate outwards
from the ILR.  At the edge of the forbidden region around the CR these waves
transfer to the long-leading branch and propagate back towards the ILR. As
the waves approach the ILR, $k$ decreases and the validity of the
tight-winding approximation becomes questionable. If it remains valid, the
waves reflect off the ILR into long-trailing waves, which propagate out
towards the CR. At the edge of the CR's forbidden region the waves morph into
short-trailing waves, which propagate back towards the ILR. As they approach
the ILR $k$ is predicted to grow without limit. In reality the wave is
absorbed as it approaches the ILR and its energy dissipates as heat. Thus the
tight-winding approximation predicts that short-leading waves gradually wind
up into short-trailing waves, which heat the disc in the vicinity of the ILR.
The unwinding of leading waves and winding-up of trailing waves is similar to
what differential rotation would do to material arms.

Similarly, the dispersion relation implies that short leading waves will
propagate inwards from the OLR to the outer edge of the forbidden region
around the CR, where they will transfer to the long branch of the dispersion
relation and move back out as long leading waves. If the tight-winding
approximation remains valid as they approach the  OLR, they will morph into
long trailing waves that propagate back inwards towards the CR and then
return as short trailing waves that eventually thermalise at the OLR.

As waves transfer from leading to trailing form near a Lindblad resonance,
the waves morph from elongated ridges of overdensity to compact blobs of
over-density.  If Toomre's
\[
Q\equiv{\sigma\kappa\over3.36 G\Sigma_0}
\]
 is small enough, self-gravity imparts a sharp inward impulse to these
blobs, so their density begins to rise. Simultaneously, differential
rotation is shearing them into trailing waves, which propagate away from the
resonance.
Consequently, the waves that reach the forbidden zone around the CR have
larger amplitude than the waves that left this zone earlier. We say the waves
have been \deffn{swing amplified}. Our formulae do not predict this behaviour
because they rely on the tight-winding approximation, which is invalid at the
crucial moment.

The gain  of the swing amplifier is a sensitive function of $Q$ and the
parameter
\[
X\equiv {k_{\rm crit}R\over m},
\]
 where $k_{\rm crit}$ is defined by equation (\ref{eq:LSK}). The smaller $X$
is, the more invalid the tight-winding approximation, and the smaller $Q$ is,
the cooler the disc.  The disc is stable to axisymmetric disturbances only if
$Q>1$.\footnote{We can show this from equation (\ref{eq:LSK}) by setting
$m=0$.} Swing amplification by a factor $>10$ is possible for $Q\lta1.5$ and
$X\lta3$.

Since a key phase in the life-cycle of waves that we just described, the
waves are unlikely to satisfy the tight-winding approximation, it is natural
to ask about solutions of the fundamental equation for normal modes,
eq.~(\ref{eq:normalm}). \cite{Toomre81} reported results obtained in this way
by his student T.~Zang. These showed that when the growth rate of a mode is
not large, the mode looks like an interference pattern between leading and
trailing waves that differ only in their amplitude, the trailing wave having
the larger amplitude. The larger the mode's growth rate is, the more the
trailing waves dominates. This finding is consistent with swing amplification
taking place as disturbances morph from leading to trailing. Another finding
was that the modes are essentially confined to the region between the ILR and
the CR. \cite{ReadE} solved for the normal modes of discs with power-law
gravitatinal potentials. They also found that the amplitudes of modes are
largest between the ILR and the CR. Their models had ``cut-out'' discs, that
is discs whose surface density tapered to negligible values at both very
small and very large radii. The growth rate of a mode depended strongly on
whether the ILR lay inside the inner cut-out, that is in a region of low
density. In this case inward propagating short-trailing waves can reflect off
the inner edge of the disc into leading waves, thus closing a feedback loop,
rather than being absorbed at the ILR. Thus solutions to our mode equation
(\ref{eq:normalm}) lend support to the qualitative understanding of spiral
structure provided by the dispersion relation for tightly-wound waves.

The bottom line is that stellar discs are responsive dynamical systems
because they support waves that can be amplified by self gravity as they move
through the disc. The degree of amplification, and therefore the disc's
responsiveness, increases sharply as the velocity dispersion falls towards
the critical value at which $Q=1$ and the disc becomes unstable to radial
fragmentation.  Much of the energy carried by the waves is thermalised in the
vicinity of a Lindblad resonance. Thus the waves heat the disc and render it
less responsive.  

\subsection{Spiral structure and normal modes}

\cite{LinShu} hypothesised that spiral structure is a manifestation of a
mildly unstable normal model of the stellar disc: they envisaged the
amplitude of this mode stabilising at a finite value as a result of energy
dissipation in interstellar gas. They developed the theory of density waves
in the expectation that the normal modes of discs could be understood in
terms of waves trapped between barriers in the same way that we picture the
modes of a laser as standing waves trapped between the laser's end mirrors.
It's now clear that the very influential Lin--Shu paradigm is based on a
misunderstanding of disc dynamics.  Waves in a stellar disc are heavily
damped already at the level of stellar dynamics because they heat the disc at
the Lindblad resonances.

From a certain perspective the failure of the Lin--Shu hypothesis is
perplexing: stellar dynamics is governed by the coupled Poisson and
collisionless Boltzmann equations. These equations are time-translation
invariant, so on group-theoretic grounds their linearised forms must have a
complete set of solutions with time dependence $\e^{\nu t}$, with $\nu$
possibly complex. Unless the system is completely stable, the evolution from
any initial condition will be dominated by the most rapidly growing normal
mode. Hence the observations must reflect such modes.

The problem with this argument is that it assumes that any initial
configuration can be represented by a superposition of normal modes. In other
words, the normal modes are assumed to be complete. The solutions to our
normal-mode equation (\ref{eq:normalm}) are not complete because in deriving
it we have used defective logic: equation (\ref{eq:fnfromPhin}) is obtained
by dividing both sides of
 \[\label{eq:vanK}
(\vn\cdot\vOmega-\omega)\delta f_\vn=\vn\cdot{\p f_0\over\p\vJ}\delta\Phi_\vn
\]
 by $\vn\cdot\vOmega-\omega$. This operation is legitimate only if
$\vn\cdot\vOmega-\omega\ne0$. If we want our normal modes to be complete, we
have to include the case $\vn\cdot\vOmega-\omega=0$ and replace equation
(\ref{eq:fnfromPhin}) by
 \[
\delta f_\vn(\vJ)={\vn\cdot\p
f_0/\p\vJ\over\vn\cdot\vOmega-\omega}\delta\Phi_\vn(\vJ)
+c_\vn(\vJ)\delta(\vn\cdot\vOmega-\omega).
\]
 In the simpler but closely analogous case of an electrostatic plasma,
 \cite{vanKampen} was able to show that by considering such {\it singular\/}
DFs Poisson's equation can be satisfied for any real value of $\omega$. In
this way we obtain a much richer set of solutions than can be obtained from
equation (\ref{eq:fnfromPhin}). All these \deffn{van Kampen modes} are
stable, and they prove to be complete, whereas the solutions we would obtain
from equation (\ref{eq:fnfromPhin}) are incomplete and as such do not form a
basis for a discussion of stability.

\subsection{Driving spiral structure}

By counting faint stars in the outer reaches of both our Galaxy and the
Andromeda nebula, M31, it has been shown that the outer parts of galaxies are
a mass of stellar streams and full of faint satellite galaxies
\citep{McConnachie,Belokurov,Bell}. From studies of the internal dynamics of
satellite galaxies, we know that these systems are heavily dominated by dark
matter, so we must anticipate that the dark-matter distribution that
surrounds a galaxy like the Milky Way is lumpy. When a lump of dark matter
sweeps through pericentre, its tidal field will launch a wave into the host
galaxy's disc, which we know to be a responsive system. The classic example
of this process in M51, which has a satellite galaxy, NGC\,5195, near the end
of one of its exceptionally strong spiral arms. Few galaxies have such a
luminous satellite so near to them, so \deffn{grand-design} spirals like that
of M51 are not prevalent.  Most galaxies will be responding simultaneously to
more than one much weaker stimulus, with the result that their spirals are
both weaker and rather chaotic.

A majority of spiral galaxies have bars at their centres. The figures of bars
are known to rotate quite rapidly in that the CR of the bar's pattern lies at
a radius that is $\sim1.2$ times the bar's length. The rotating gravitational
field of the bar must perturb the disc, and from the discussion of
\S\ref{sec:tightwind} we would expect the surrounding disc to show spiral
structure at the pattern speed of the bar that extends from near the end of
the bar to the OLR.  However, both N-body simulations and observations show
that the disc's principal response is at a lower pattern speed than that of
the bar \citep{SparkeS}, so bar excites spiral structure that lies inside its CR. This
finding is consistent with the tendency of the solutions to the normal-mode
equation (\ref{eq:normalm}) to have significant amplitudes only inside CR.
Crucially it implies that the response to the bar rotates more slowly than
the bar, so the relative phases of the features is constantly changing.

\section{Conclusion}

The key approximation of stellar dynamics is that stellar systems are
collisionless, so the actual motion is well approximated by motion in a
smooth potential. Orbits in smooth potentials are mostly quasiperiodic, and
when they are not, it is possible to construct a nearby
Hamiltonian in which the same initial conditions yield quasiperiodic motion.
Therefore quasiperiodic motion is an excellent starting point for stellar
dynamics.

Quasiperiodic orbits display an elegant structure that is captured by
angle-action coordinates: each orbit is a three-torus. The angle variables
specify where on its torus a star is, and they evolve linearly in time. If
the frequencies are incommensurable (as they nearly always should be) a 
star's probability density should be independent of angle variables, so the
distribution function depends only on the actions. The actions
provide a geometrical quantification of the orbit.

Normally the true Hamiltonian will differ from the approximate one that
admits angle-action variables. The small difference $h$ between the true
Hamiltonian and the approximate one can be important if it generates forces
that act in the same way over extended periods of time. In the vicinity of
resonances this may happen, and then $h$ may change the dynamics
qualitatively. The overall impact on the dynamics of a galaxy is nevertheless
likely to be small if the resonance is isolated. When several resonances are
simultaneously active, however, chaos can be generated, and stars may slowly
diffuse through phase space. This process is likely to be important for the
secular evolution of barred galaxies, but our understanding of it is
currently inadequate.

The potential of a real galaxy is always fluctuating, and fluctuations are of
fundamental importance because they alone permit stars to exchange energy.
No matter what their physical origin, fluctuations will drive the system
towards unattainable thermodynamic equilibrium, especially by enhancing
core-halo structure.  Two-body relaxation is mostly due to Poisson
fluctuations in the number of stars in large volumes, and is an exceedingly
slow process in galaxies. Hence in galactic dynamics we focus on fluctuations
due to the motion of massive bodies (giant molecular clouds, spiral arms
dark-halo lumps, star-clusters and satellite systems). Fluctuations cause
stars to diffuse through action space, and this diffusion is  observed
in the solar neighbourhood. The diffusion coefficients can be calculated from
the temporal power spectrum of the fluctuations or empirically determined
from observations of solar-neighbourhood stars. Theory and observation are
reasonably consistent, but there is plenty of scope for tightening
constraints.

Spiral structure is an important source of fluctuations. Its dominant effect
is the creation of transient resonances, which by first trapping and then
releasing stars cause them to move from inside the corotation circle
outwards, and vice versa. The random velocities of stars are not increased by
such churning, but stars with similar ages but different metallicities are
mixed up. In addition to churning the disc around corotation, spiral
structure moves angular momentum outwards, from ILR to OLR, in the process
heating the disc in the vicinity of the Lindblad resonances, especially the
ILR. 

Spiral structure is not fully understood because it is an inherently
non-linear and global phenomenon. Lin-Shu-Kalnajs density-wave theory is
restricted to the linear case and assumes tightly-would arms to make the
physics essentially local. It predicts that tightly-wound leading waves
propagate through a portion of the disc, unwinding as they go, so they
inevitably violate the tight-winding approximation. As the waves pass from
leading to trailing form, they are amplified by a process that lies beyond
linear theory, and eventually their energy is thermalised at a Lindblad
resonance by an analogue of Landau damping. The inaccuracy inherent in using
the tight-winding approximation can be eliminated by solving the exact
equation for normal modes. Such solutions confirm the basic picture derived
with the tight-winding approximation but reveals a preference of spiral
structure to lie inside CR. Unfortunately, van Kampen's work on electrostatic
plasmas implies that the solutions of the normal-mode equation are not
complete, so they do not provide a secure basis for understanding the
dynamics of discs.

Self-gravitating discs are responsive systems because any disturbance is
liable to excite leading waves, which may amplify significantly as they
morph into trailing waves. Since the amplification becomes weaker as the disc
heats, a pure stellar disc becomes less responsive as its velocity dispersion
rises. Gas is an essential ingredient of a spiral galaxy because (i) it
dissipates the energy of spiral waves, (ii) through star formation it
constantly replenishes the population of stars with low velocity dispersion
as spiral structure increases the velocity dispersion of older stars, and
(iii) it makes any spiral gravitational potential observationally conspicuous
by forming  dust lanes and luminous blue stars near its troughs.

Galaxies live in the noisy environments of their dark halos, into which
clumps with various masses are continually falling. As they pass through
pericentre such lumps may excite spiral structure. Any spiral structure will
quickly heat the disc. If the disc is relatively cold and therefore
responsive, much more energy will be converted into heat than was imparted by
the exciting lump: by shifting angular momentum out through the disc, spiral
structure makes gravitational energy available for random motions, and thus
increases the disc's entropy. It is important never to lose sight of the fact
that a disc of stars on nearly circular orbits is occupying only a tiny
fraction of the phase space that is energetically accessible to it. Any
random process will scatter its stars into a broader distribution in phase
space, and thus make it a hotter, thicker disc.

%


\begin{thebibliography}{}

\bibitem[Arnold(1978)]{Arnold}
Arnold V.I., 1978, {\it Mathematical Methods of Classical Mechanics},
Springer: New York

\bibitem[Aumer \& Binney(2009)]{AumerB}
Aumer M., Binney J.J., 2009, MNRAS, 397, 1286

\bibitem[Bell et al.(2008)]{Bell}
Bell E.F., Zucker D.B., et al.~2008, ApJ, 680, 295

\bibitem[Belokurov et al.(2006)]{Belokurov}
Belokurov, V., Zucker D.B., et al., 2006, ApJ, 642, 137

\bibitem[Binney(1982)]{Binney82}
Binney J., 1982, MNRAS, 201, 1 

\bibitem[Binney(2010)]{Binney10}
Binney J., 2010, MNRAS, 210, 2318

\bibitem[Binney \& Tremaine(2008)]{BT08}
Binney J., Tremaine S., 2008, {\it Galactic Dynamics}, Princeton University
Press: Princeton

\bibitem[Binney \& Lacey(1988)]{BinneyL}
Binney J., Lacey C., 1988, MNRAS, 230, 597

\bibitem[Gradshteyn \& Ryzhik(1965)]{GradstynR}
Gradshteyn I.S., Ryzhik I.M., 1965, {\it Tables of Integrals, Series and
Products}, Academic Press: New York

\bibitem[Kaasalainen \& Binney(1994)]{KaasalainenB94}
Kaasalainen M., Binney J., 1994, PhRvL, 73, 2377

\bibitem[Kalnajs(1977)]{Kalnajs}
Kalnajs A., 1977, ApJ, 212, 637

\bibitem[Lin \& Shu(1966)]{LinShu}
Lin C.C., Shu F.H., 1966, Proc.\ Nat.\ Sci., 55, 229

\bibitem[McConnachie et al.(2009)]{McConnachie}
McConnachie A.W., Irwin M.J., et al., 2009, Nature, 461, 66

\bibitem[Merritt \& Valluri(1999)]{ValluriM}
Merritt D., Valluri M., 1999, AJ, 118, 1177

\bibitem[Read \& Evans(1998)]{ReadE}
Read J., Evans N.W., 1998, MNRAS, 300, 106

\bibitem[Rix \& Zarisky(1995)]{RixZ}
Rix H.-W., Zaritsky D., 1995, ApJ, 447, 82

\bibitem[Sparke \& Sellwood(1988)]{SparkeS}
Sparke L.S., Sellwood J.A., 1988, MNRAS, 231, 25

\bibitem[Sridhar \& Touma(1996)]{Touma}
Sridhar S., Touma J., 1996, MNRAS, 279, 1263

\bibitem[Sellwood(2008)]{Sellwood}
Sellwood J.A., 2008, arXiv0803.1574

\bibitem[Sellwood \& Binney(2002)]{SellwoodB}
Sellwood J.A., Binney J.J., 2002, MNRAS, 336, 785

\bibitem[Sellwood \& Kahn(1991)]{SellwoodK}
Sellwood J.A., Kahn F.D., 1991, MNRAS, 250, 278

\bibitem[Spitzer \& Scwarzschild(1953)]{SpitzerS}
Spitzer L., Schwarzschild M., 1953, ApJ, 118, 106

\bibitem[Toomre(1964)]{Toomre64}
Toomre A., 1964, ApJ, 139, 1217

\bibitem[Toomre(1969)]{Toomre69}
Toomre A., 1969, ApJ, 158, 899

\bibitem[Toomre(1981)]{Toomre81}
Toomre A., 1981, in {\it The Structure and Evolution of Normal Galaxies},
ed.~S.M. Fall \& D. Lynden-Bell, Cambridge University Press: Cambridge,
p.~111

\bibitem[van Kampen(1955)]{vanKampen}
van Kampen N.G., 1955, Physica, 21, 949

\bibitem[Wielen(1977)]{Wielen}
Wielen R., 1977, A\&A, 60, 263


\end{thebibliography}
\end{document}